\documentclass{aa}  
\usepackage[switch]{lineno}
\usepackage{xcolor}
\usepackage{graphicx}
\usepackage{txfonts}
\usepackage[T1]{fontenc}
\usepackage{comment}
\usepackage{orcidlink}
\usepackage{hyperref}
\hypersetup{
  colorlinks,
  citecolor=blue,
  linkcolor=blue,
  urlcolor=blue}

\begin{document} 

      \title{He-star donor pathway for the hypervelocity star D6–2}

   \titlerunning{He-star donor pathway for the hypervelocity star D6–2}

   \author{Abinaya Swaruba Rajamuthukumar\inst{1}\orcidlink{0000-0002-1872-0124}
         \and
        R\"udiger Pakmor\inst{1}\orcidlink{0000-0003-3308-2420}
        \and
        Stephen Justham\inst{1}\orcidlink{0000-0001-7969-1569}
        \and
        Aakash Bhat\inst{2}\orcidlink{0000-0002-4803-5902}
        \and
        Ken J. Shen\inst{3}\orcidlink{0000-0002-9632-6106}
        }
        
\institute{Max-Planck-Institut für Astrophysik, Karl-Schwarzschild-Straße 1, 85748 Garching bei München, Germany\\
              \email{abinaya@mpa-garching.mpg.de}
            \and
             Institut f\"{u}r Physik und Astronomie, Universit\"{a}t Potsdam, Haus 28, Karl-Liebknecht-Str. 24/25, 14476 Potsdam-Golm, Germany
            \and
            Department of Astronomy and Theoretical Astrophysics Center, University of California, Berkeley, CA 94720-3411, USA
            }
    \authorrunning{A.S.Rajamuthukumar et al}

  \abstract
 {Type~Ia supernovae are thermonuclear explosions of white dwarfs, yet the nature of their progenitor systems remains uncertain. Recent discoveries of hypervelocity stars provide unique constraints, as these stars likely represent the surviving companions of such explosions. Using detailed binary evolution models computed with \texttt{MESA} and population synthesis with \texttt{MSE}, we investigate the outcomes of hot subdwarf + white dwarf binaries undergoing helium accretion. We find that donors can nearly exhaust their helium and form compact, C/O cores before explosion. The predicted ejection velocities span a broad distribution reaching up to $\sim 1000\,\mathrm{km\,s^{-1}}$, with D6-2 representing the extreme high-velocity tail of this population. We estimate analytically that the thin residual helium envelope can be stripped by the supernova ejecta, producing a C/O-rich surface composition consistent with the observed spectrum.
 The Type~Ia supernova rate from this channel is ${\sim}(1.69\pm0.06)\times10^{-5}\,\rm M_\odot^{-1}$, consistent with 1\% of the observed Type Ia supernova rate. Hot subdwarf + white dwarf binaries containing nearly exhausted He-star donors can therefore naturally explain the velocity and composition of D6-2 while providing a quantitatively consistent contribution to the observed Type~Ia supernova rate. Our models predict a distribution of surviving donor remnants with various core He fractions and with ejection velocities   extending down to $\sim 450\,\mathrm{ km\,s^{-1}}$. The orbital velocities of donor stars in this progenitor channel naturally yield orbital velocities consistent with US 708, LP 40–365 stars, and D6‑2, indicating that a single class of thermonuclear supernova progenitors can account for their entire range of ejection velocities.
}


   \maketitle
%
  {}

%

\section{Introduction}
\label{sec:intro}

Type Ia supernovae are among the most energetic astrophysical transients and are of fundamental importance. They serve as standardizable candles for cosmology, having led to the discovery of the accelerating expansion of the universe \citep{1998AJ....116.1009R,1998ApJ...507...46S,1999ApJ...517..565P} and they are also major sources of iron group elements that drive the chemical evolution of galaxies \citep{1969ApJ...157..623C,1982ApJ...253..785A,2013FrPhy...8..116H}. Despite decades of theoretical and observational efforts, the nature of their progenitor systems remains uncertain (for recent reviews, see \citealt{2023RAA....23h2001L,2025A&ARv..33....1R}). The prevailing view is that these events arise from the thermonuclear explosion of a white dwarf in a binary system. Two commonly discussed progenitor pathways exist: in the single-degenerate channel, the companion is a non-degenerate star \citep{1973ApJ...186.1007W,1982ApJ...253..798N,1982ApJ...257..780N,1984ApJ...286..644N}, whereas in the double-degenerate channel, the companion is another white dwarf \citep{1984ApJ...277..355W,1984ApJ...284..719I}. Direct observations have so far failed to reveal any survivors in supernova remnants \citep{2009ApJ...701.1665K, 2012Natur.481..164S, 2018ApJ...862..124R,2023ApJ...950L..10S}. However, growing detections of hypervelocity runaway stars may provide indirect clues to their origin \citep{2005A&A...444L..61H,2015Sci...347.1126G,2015ApJ...804...49B,2018ApJ...865...15S,2023OJAp....6E..28E,2025MNRAS.541.2231H}.
\begin{figure*}
    \centering
    \includegraphics[trim = 10 0 50 0 ,width=1\textwidth]{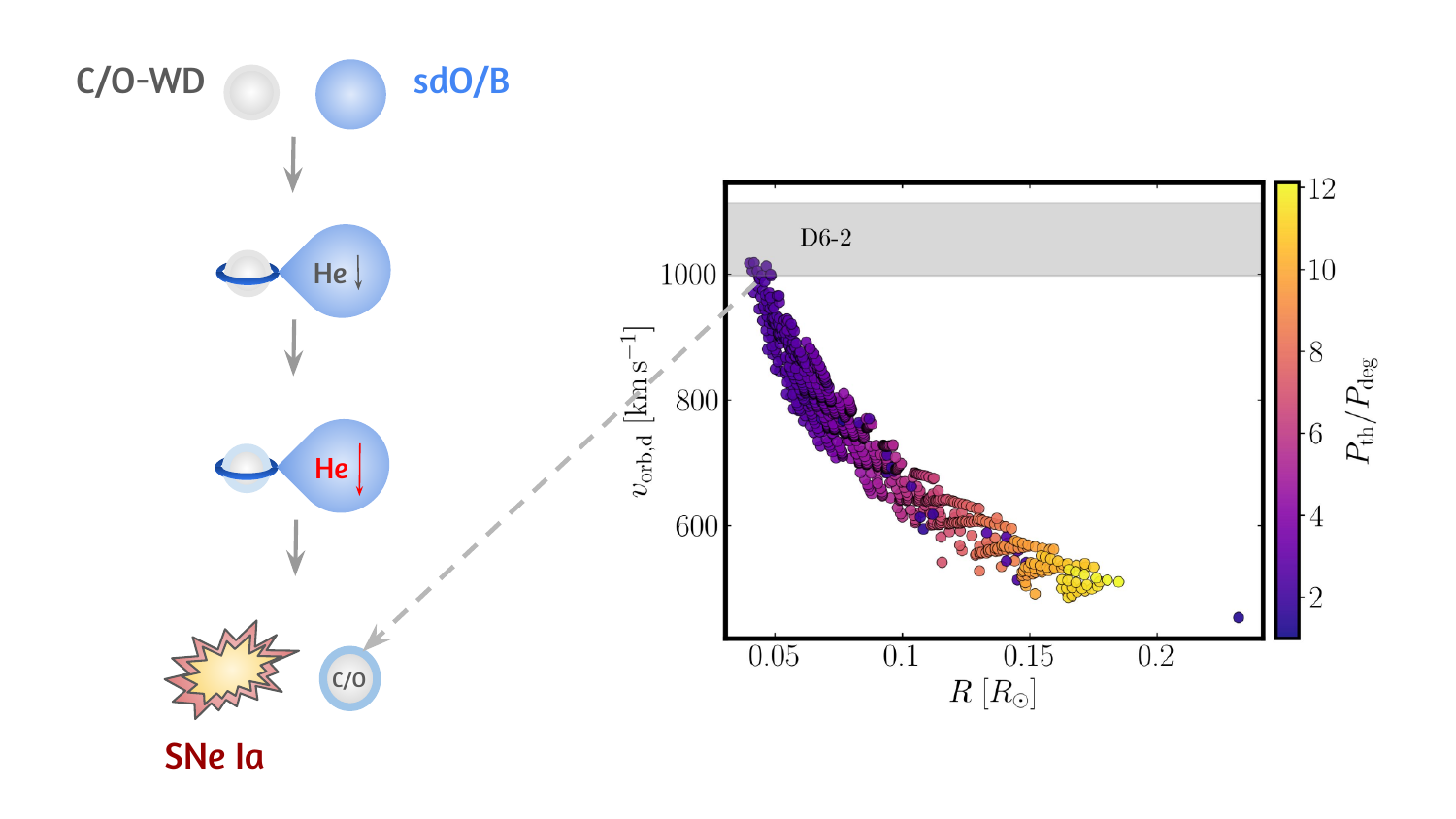}
    \caption{Ejection properties of donors in hot subdwarf + white dwarf binaries. 
Left: Schematic illustrating the formation of C/O core donor stars with varying degrees of degeneracy in CO white dwarf + hot subdwarf binaries. 
Right: Radius versus orbital velocity of the donors at the time of accretor explosion, with the color bar indicating the degree of degeneracy at the time of explosion. 
The grey region shows the velocity range observed for D6-2. 
Some of our models reach this velocity range, demonstrating that evolved donors in hot subdwarf + white dwarf systems can naturally reproduce the kinematic properties of D6-2.}
\end{figure*}\label{fig:schematic} The hypervelocity He-rich star US 708 \citep{2005A&A...444L..61H} has been explained as the runaway donor star from a thermonuclear explosion \citep{2009A&A...493.1081J, 2015Sci...347.1126G, 2015ApJ...804...49B,2022A&A...663A..91N}.   Several hypervelocity stars have also been identified with Gaia \citep{2018ApJ...865...15S,2023OJAp....6E..28E} which, despite having distinctly different appearances from US 708, also seem very likely to be the surviving companions of thermonuclear explosions, flung out at roughly their pre-explosion orbital velocities. These hypervelocity stars thus represent perhaps the closest observational evidence for Type Ia supernova progenitor channels.

The velocity of these hypervelocity stars encodes information on their pre-explosion compactness. Systems with main-sequence donors typically yield runaway velocities of a few hundred~$\mathrm{km\,s^{-1}}$ \citep{2008ApJ...677L.109H}, helium (He) star donors can reach up to $\simeq 1000~\mathrm{km\,s^{-1}}$ \citep{2020A&A...641A..52N,2024arXiv241108099R}, and ultra-compact double white dwarf systems can eject survivors at velocities greater than $1000~\mathrm{km\,s^{-1}}$ \citep{2018ApJ...865...15S,2022MNRAS.512.6122C,2025NatAs.tmp..168G,2025arXiv251011781P,2025arXiv251012197B}. Intriguingly, the velocity of D6-2, reported by \cite{2018ApJ...865...15S}, lies at the boundary of the expected ranges for He-star and double white dwarf progenitors. Since its velocity is too low to arise from a double CO white dwarf system, previous studies \citep[e.g.][]{2024ApJ...973...65W,2025arXiv250812529W} have suggested that it could originate from a He white dwarf donor. While most He WD donor models remain inflated and luminous for longer than D6-2’s observed compact and faint state, the lowest-mass donors ($\approx 0.08 - 0.10 \,\mathrm{M_\odot}$) can evolve to a compact, faint remnant on timescales comparable to its inferred age.

One plausible evolutionary pathway that can naturally bridge this gap is the high velocity runaway stars from He accreting white dwarfs with non-degenerate donors \citep{2019ApJ...887...68B,2020A&A...641A..52N,2021A&A...646L...8N,2022ApJ...925L..12K,2024MNRAS.527.2072D}. In such systems, the hot subdwarf, initially a core He burning star, can evolve into a  He-star with decreased core He fraction or a low mass CO white dwarf with a thick He shell by the time the companion white dwarf undergoes a He-shell detonation. The explosion then unbinds the system, ejecting the evolved donor at velocities consistent with its pre-explosion orbital velocities. In our previous work \citep{2024arXiv241108099R}, we modeled the binary evolution of hot subdwarf and white dwarf systems, identifying the conditions under which mass transfer triggers thermonuclear explosions. We also estimated that velocities of these donors can reach up to $1000~\mathrm{km\,s^{-1}}$. In this paper, we expand \cite{2024arXiv241108099R} by combining detailed MESA binary evolution models with population synthesis calculations to address three key questions: 1) Can He-star donor channels explain hypervelocity runaway white dwarfs? 2) Can we reproduce the observed properties of D6-2? 3) What are the expected rates of such hypervelocity runaway stars?

The paper is structured as follows. In Section 2, we describe our methods. Section 3 presents velocities from MESA simulations and their connection to D6-2. In Section 4, we outline the Type Ia supernova rate and velocity distributions. We discuss our results in Section 5 and conclude in Section 6.

\section{Methods}
\label{sec:methods}

We adopt the same numerical setup and physical assumptions as described in detail in \citet{2024arXiv241108099R}. Here, we briefly summarize the key aspects for completeness.

All binary evolution models are computed using \texttt{Modules for Experiments in Stellar Astrophysics} (MESA, version r23.05.1; \citealt{2011ApJS..192....3P,2013ApJS..208....4P,2015ApJS..220...15P,2018ApJS..234...34P,2019ApJS..243...10P, 2023ApJS..265...15J}), following the evolution of hot subdwarf + CO white dwarf binaries. The models include convective mixing treated using time-dependent convection (\texttt{MLT\_option = 'TDC'}) with $\alpha_{\mathrm{MLT}} = 2$, and employ the \texttt{Schwarzschild} criterion for convective instability. Nuclear networks include \texttt{basic\_plus\_fe56\_ni58.net} for the hot subdwarf and a modified \texttt{nco.net} \citep{2017ApJ...845...97B} for the white dwarf, capturing the NCO chain capable of triggering He-shell ignition at densities $\gtrsim10^6\,\mathrm{g\,cm^{-3}}$. Mass loss on the RGB and AGB is treated with Reimers’ and Blöcker’s prescriptions, while winds are neglected during the hot subdwarf phase.  

The hot subdwarf donors are constructed by stripping red giant progenitors after core He ignition, leaving a residual hydrogen envelope mass of $3 \times 10^{-4}\, \rm M_\odot$. The CO white dwarfs are produced using a modified \texttt{make\_co\_wd} suite, cooled to $3\times10^8\, \rm yr$. The donor masses span $0.33$ – $0.8\,\rm M_\odot$, and the CO white dwarf accretors range from $0.7$–$1.1\,\rm M_\odot$. The binaries are initialized with orbital periods between 36 minutes and 7.2 hours, logarithmically spaced to capture systems that undergo mass transfer at various donor evolutionary stages.

We assume conservative mass transfer, and compute mass transfer rate using MESA’s Kolb prescription \citep{1990A&A...236..385K, 2015ApJS..220...15P}. To avoid computationally expensive H novae cycles, the accreted material is assumed to be He rich throughout the evolution, implemented via a modified \texttt{binary\_mdot.f90} (available in the associated Zenodo repository\footnote{\url{https://zenodo.org/records/13473758}}). The accretor’s thermal and structural response, including compressional heating, is fully treated within MESA’s energy equations. All models assume solar metallicity ($Z=0.02$), no overshooting, and no gravitational settling.

Each system is evolved until one of the following occurs: (1) He ignition in the accretor reaches a luminosity of $5 \times 10^{4}\,\rm L_\odot$, or (2) the donor becomes a white dwarf, resulting in a detached double white dwarf system. Following \citet{1994ApJ...423..371W}, we classify ignitions as detonations if the He ignition occurs at densities $\rho \ge 10^{6}\,\mathrm{g\,cm^{-3}}$. Achieving detonation densities requires a massive accreted He shell ($>0.05\,\rm M_\odot$; \citealt{2024arXiv241108099R}). While we do not evolve beyond the He-shell detonation, we assume that the initial He-shell detonation triggers a secondary C/O core detonation. Recent simulations have shown that even comparatively low-mass He shells are capable of igniting the underlying core \citep{2024ApJ...975..127S}. We assume that the donor’s ejection velocity following the explosion is approximately given by its orbital velocity at the time of ignition. While the ejection velocities are generally similar to the pre-explosion orbital velocities, small differences may arise from the impact process \citep{2019ApJ...887...68B,2024OJAp....7E...7B,2025A&A...693A.114B,2025arXiv251011781P,2025arXiv251012197B}. We plan to perform detailed 3D hydrodynamical impact simulations to explore the explosion dynamics in greater detail, which will also allow us to determine the ejection velocities more precisely.

\begin{figure}
    \centering
    \includegraphics[trim = 0 0 0 0 ,width=0.5\textwidth]{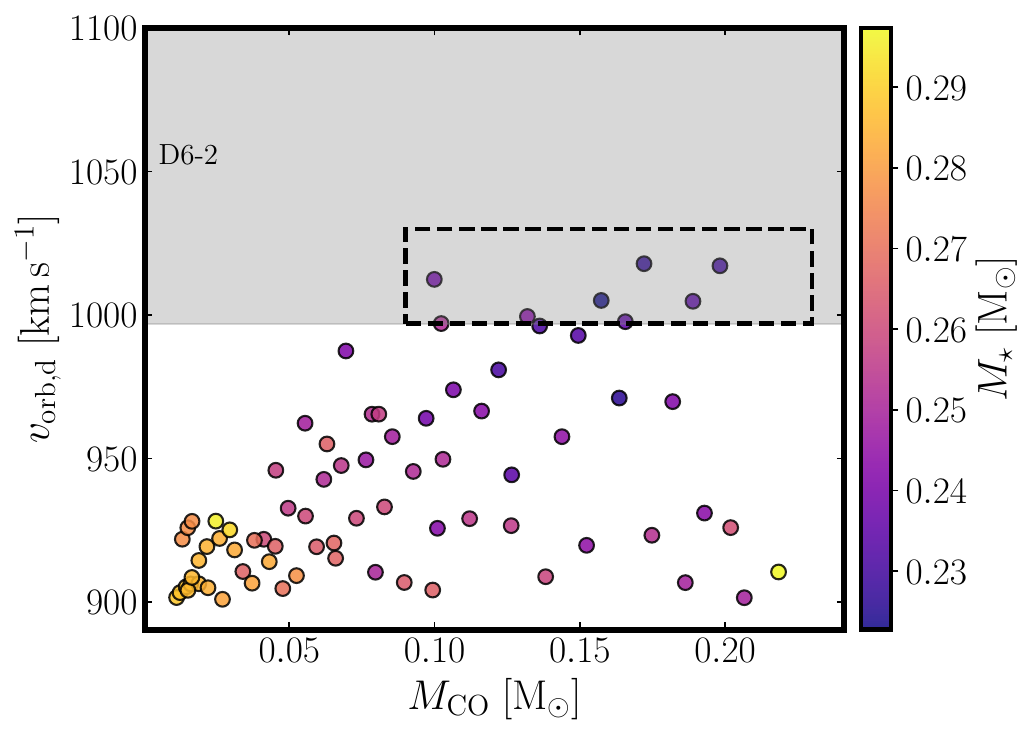}
    \caption{Ejection velocity of donor stars as a function of their C/O core mass at the time of the companion’s explosion. The color bar indicates the total donor mass. The shaded region marks the measured velocity of D6-2. Models within this region possess C/O core masses of $\sim 0.1 - 0.2 \,\rm M_\odot$, consistent with the observed properties of D6-2.}
    \label{fig:Vorb_MCO}
\end{figure}
\begin{figure}
    \centering
     \includegraphics[width=1\columnwidth]{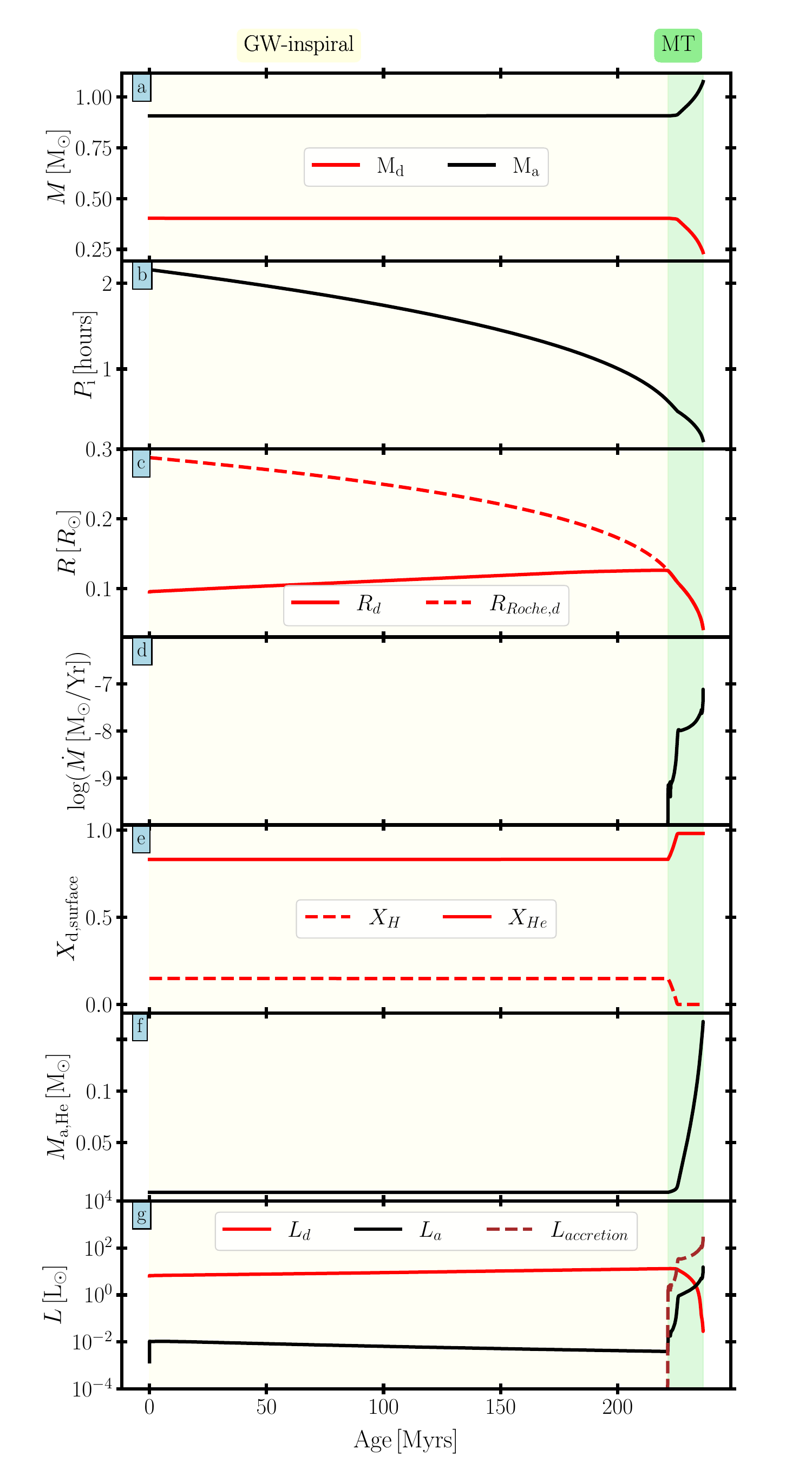}
    \caption{An example of a binary with $\mathrm{M_d = 0.4\,\mathrm{M_{\odot}}}$, $\mathrm{M_a = 0.9 \,\mathrm{M_{\odot}}}$, and $\mathrm{P_i = 2.16\, hours}$,  where a thermonuclear explosion occurs in the white dwarf (accretor). The background colors yellow and green represent the gravitational-wave inspiral phase (GW-inspiral) and the mass transfer phase (MT), respectively.  Panels (a) through (f) show the evolution of various parameters: (a) \textbf{Mass Evolution:} $M_a$ (accretor) and $M_d$ (donor); (b) \textbf{Orbital Period Evolution}; (c) \textbf{Radius Evolution:} $R_d$ (donor's radius) and $R_{roche,d}$ (Roche radius of the donor); (d) \textbf{mass transfer rate}; (e) \textbf{Evolution of surface mass fraction:} $X_H$ (hydrogen) and $X_{He}$ (He) in the donor; (f) \textbf{He Mass on the Accretor}; (g) \textbf{Luminosity Evolution:} $L_a$ (accretor), $L_d$ (donor), and $L_{\text{accretion}}$ (accretion). The accretor gains mass at the rate of $\sim 10^{-8} \, \mathrm{M_{\odot}yr^{-1}}$, resulting in the He ignition in the He layers denser than the assumed critical density for detonation ($> 10^{6} \, \mathrm{g\,cm^{-3}}$). }
    \label{fig:Example}
\end{figure}

\begin{figure*}
    \centering
    \includegraphics[trim = 0 0 0 0 ,width=1\textwidth]{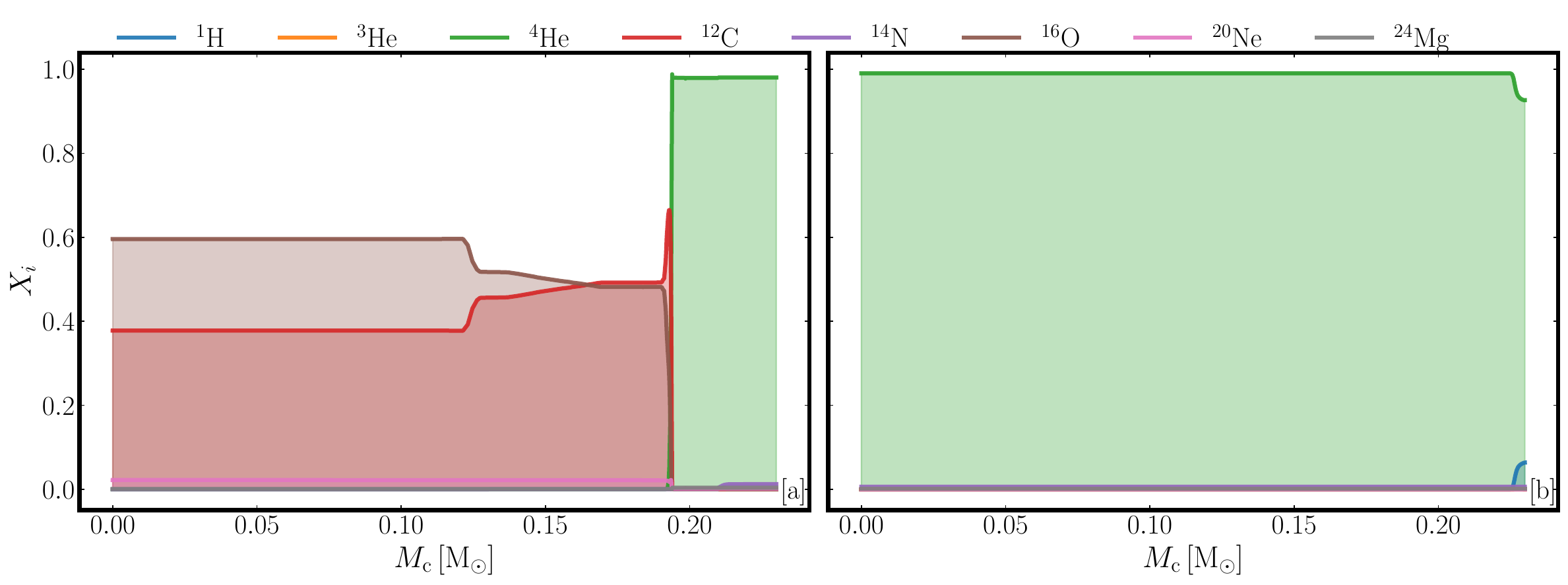}
    \caption{Internal composition profiles of the donor star at the time of explosion (panel a) compared to a canonical He white dwarf (panel b). The x-axis shows the enclosed mass coordinate, and the y-axis indicates mass fractions of key isotopes. The hot subdwarf donor has already converted $\gtrsim 90 \%$ of its He into carbon and oxygen, in contrast to the He-dominated composition of the He white dwarf model.}
    \label{fig:composition}
\end{figure*}

\section{Runaway velocities and donor evolution}

Our MESA binary models predict runaway velocities ranging from
$\sim\,450$ to $\sim\, 1020\,\mathrm{km\,s^{-1}}$, increasing systematically with donor
compactness (Fig.~\ref{fig:schematic}).  
The color scale in Fig.~\ref{fig:schematic} represents the ratio of thermal to
degeneracy pressure in the center, ($P_{\mathrm{th}}/P_{\mathrm{deg}}$), which serves as a proxy for the degree of degeneracy.  
Systems with larger orbital velocities correspond to donors with smaller radii and lower $P_{\mathrm{th}}/P_{\mathrm{deg}}$ values, indicating that some of the fastest objects are nearly degenerate at the time of explosion.  
Donors that achieve $v_{\mathrm{orb}}\gtrsim1000\,\mathrm{km\,s^{-1}}$
possess compact interiors dominated by carbon and oxygen.

Although the binaries initially start with core He burning hot subdwarfs, many donors continue to evolve during extended phases of stable mass transfer, lasting up to several tens to hundreds of Myr. During this period, He burning proceeds nearly to completion in the core, transforming the donor into a transitioning white dwarf with a central C/O composition. The resulting stars are compact, with a higher degree of degeneracy, and surrounded by only a low-mass He envelope. These stars will further cool to become a fully degenerate white dwarfs. Despite starting mass transfer in a single degenerate system, the donor might not be clearly non-degenerate anymore at the time of explosion. Fig.~\ref{fig:Vorb_MCO} shows the correlation between the donor’s C/O core mass and its ejection velocity, highlighting that this channel can reach the fastest velocity with a far-evolved core. The shaded band marks the observed velocity range of D6-2 \citep{2023OJAp....6E..28E}, illustrating that some donors with $M_{\mathrm{CO}}\simeq0.1$–$0.2\,\rm M_\odot$ occupy this regime.

To illustrate this evolution in detail, we show in Fig.~\ref{fig:Example} a system with an initial donor mass $M_\mathrm{d}=0.4\,\rm M_\odot$, an accretor mass $M_\mathrm{a}=0.9\,\rm M_\odot$, and an initial orbital period $P_\mathrm{i}=2.16\,\mathrm{h}$. The system first undergoes a short gravitational-wave–driven inspiral (yellow region) before stable mass transfer begins (green region).
After $\simeq227\,\mathrm{Myr}$, the donor fills its Roche lobe, having already converted $\sim90\%$ of its central He into carbon and oxygen. The core remains thermally supported by nuclear reactions and is not yet degenerate. Mass transfer proceeds at $\dot{M}\simeq10^{-8}\,\rm M_\odot\,\mathrm{yr^{-1}}$ for roughly $12\,\mathrm{Myr}$, during which the donor becomes increasingly degenerate and contracts. The accretion onto the white dwarf eventually heats enough to ignite He. Since the He fusion occurs at sufficiently higher densities  ($\rho_{\mathrm{ign}}\ge10^{6}\,\mathrm{g\,cm^{-3}}$), we assume that this He detonation then triggers ignition of the C/O core.
At this point, the donor’s core consists of $\sim90\%$ carbon and oxygen and $\lesssim10\%$ He. The disruption of the system ejects the donor at $v_{\mathrm{orb}}\simeq1000\,\mathrm{km\,s^{-1}}$, producing a fast moving, C/O dominated transitioning white dwarf.

Fig.~\ref{fig:composition} compares the internal composition of the
donor at the moment of explosion with that of a canonical He white dwarf.
The hot subdwarf + white dwarf donor exhibits a C/O dominated core
($X_{\mathrm{O}}\approx0.6$, $X_{\mathrm{C}}\approx0.4$) and only a thin, He rich surface layer, whereas the He white dwarf remains He dominated throughout.  
This compositional contrast demonstrates that some of the donors in the fastest systems have evolved beyond the He-star stage and are transitioning toward low mass CO white dwarfs at the time of the explosion.

\subsection{Connection to D6-2}

The recent discovery of hypervelocity white dwarfs by \citet{2018ApJ...865...15S} provides potential observational evidence for surviving companions of Type Ia supernova explosions. 
They identify three candidate stars with total velocities between $\sim$ 1000$\,\rm km s^{-1}$ and 3000$\,\rm km s^{-1}$, and interpret them as white dwarfs ejected at (or near) the pre-explosion orbital velocity in double white dwarf binary detonations. One of these candidates  (D6-2) even traces back to a faint supernova remnant, reinforcing the ejection hypothesis.

While the double degenerate channel is compelling for producing extreme velocity white dwarf survivors, it does not inherently predict intermediate velocity, C/O dominated survivors arising from He-star donors. In particular, models of double-degenerate systems \citep[e.g.,][]{2018ApJ...865...15S,2025ApJ...982....6S} suggest that the minimum ejection velocity is typically $\gtrsim 1000\,\mathrm{km\,s^{-1}}$, meaning that D6-2 would represent an lower end tail of the velocity distribution in the He white dwarf + CO white dwarf channel if the CO white dwarf explodes. Our models of hot subdwarf + white dwarf binaries, by contrast, predict a velocity distribution at lower velocities that extends in its high velocity tail to  $\sim 1000\,\mathrm{km\,s^{-1}}$, overlapping precisely with D6-2’s inferred velocity band (Fig.~\ref{fig:Vorb_MCO}).  
Moreover, our donors undergo extensive He burning and evolve into compact, nearly degenerate C/O core stars with thin He layers (Fig.~\ref{fig:Vorb_MCO}).  
As shown in Section~\ref{sec:he_stripping}, simple analytical estimates of the ejecta–companion interaction indicate that for such compact donors ($R_*\simeq0.03$–$0.05\,\rm R_\odot$, $a\simeq0.1$–$0.2\,\rm R_\odot$), the supernova impact can remove
$M_{\mathrm{strip}}\sim10^{-3}$–$10^{-1}\,\rm M_\odot$ of He comparable to or exceeding the donor’s remaining envelope mass.  
Hence, even if a thin He layer survives until explosion, it can plausibly be stripped by the ejecta, leaving a C/O dominated atmosphere consistent with D6-2’s observed spectrum. Nonetheless, He would not be directly observable at D6-2’s temperature, so its presence cannot be ruled out. This atmosphere could then be further enriched with additional elements deposited by the supernova impact.

Future observations and spectroscopic follow-up of hypervelocity white dwarf candidates can test this hypothesis. Hence, our model strengthens the identification of D6-2 as a distinct product of He-star donor evolution.

\subsection{Stripping of He envelope}
\label{sec:he_stripping}

We estimate the mass removed from a He companion by supernova ejecta using two complementary, order-of-magnitude limits: an \emph{energy-limited} estimate and a \emph{momentum-limited} estimate. The geometric interception of ejecta is characterized by the solid-angle fraction
\begin{equation}
f_\Omega \;=\; \frac{\pi R_*^2}{4\pi a^2} \;=\; \frac{R_*^2}{4a^2},
\end{equation}
for a companion of radius $R_*$ at orbital separation $a$.

\subsection*{Energy-limited estimate}
If a fraction $\epsilon_{\rm dep}$ of the kinetic energy intercepted by the companion is available to unbind material, the deposited energy is
\begin{equation}
E_{\rm dep} \;=\; \epsilon_{\rm dep}\, f_\Omega\, E_{\rm k,ej},
\end{equation}
where $E_{\rm k,ej}$ is the total ejecta kinetic energy. Approximating the characteristic specific binding energy near the companion surface by
\begin{equation}
\varepsilon_{\rm bind} \;\sim\; \frac{G M_*}{\lambda R_*},
\end{equation}
where $\lambda$ is a structural factor that accounts for the companion's internal mass distribution, the energy-limited stripped mass is
\begin{equation}
M_{\rm strip}^{(E)} \;\sim\; \frac{E_{\rm dep}}{\varepsilon_{\rm bind}}
\;=\; \epsilon_{\rm dep}\,\frac{E_{\rm k,ej}}{4}\,
\frac{\lambda R_*^3}{a^2 G M_*}.
\label{eq:Mstrip_energy_combined_lambda}
\end{equation}
$\epsilon_{\rm dep}\!\in\!(0,1)$ parametrizes the fraction of energy deposited that goes into stripping of He.

\subsection*{Momentum-limited estimate}
If bulk momentum transfer dominates the removal of material, the relevant budget is the intercepted ejecta momentum
\begin{equation}
p_{\rm int} \;=\; f_\Omega\, M_{\rm ej}\, v_{\rm ej},
\end{equation}
where $M_{\rm ej}$ is the ejecta mass and $v_{\rm ej}\!\equiv\!\sqrt{2E_{\rm k,ej}/M_{\rm ej}}$ a characteristic ejecta speed. Only a fraction $\epsilon_p$ of this intercepted momentum couples to accelerating donor material; the deposited momentum is $p_{\rm dep}=\epsilon_p p_{\rm int}$. Equating this to the momentum needed to accelerate stripped material to the escape speed $v_{\rm esc}\simeq\sqrt{2GM_*/R_*}$ gives
\begin{equation}
M_{\rm strip}^{(p)} \;\sim\; \frac{p_{\rm dep}}{v_{\rm esc}}
\;=\; \epsilon_p\,\frac{R_*^2}{4a^2}\, M_{\rm ej}\, \frac{v_{\rm ej}}{\sqrt{2GM_*/R_*}}.
\label{eq:Mstrip_momentum_combined}
\end{equation}

This limit describes direct momentum stripping of the outer layers, characterized by the momentum coupling efficiency $\epsilon_p$.

\subsection*{Fiducial Scalings and Numerical Estimates}

Using fiducial values appropriate for compact donors and Type Ia supernova like ejecta (and from MESA models we adopt $\lambda \simeq 0.7$), 

\begin{align*}
M_* &= 0.23\,\rm M_\odot, & R_* &= 0.04\,\rm R_\odot, & a &= 0.17\,\rm R_\odot,\\
E_{\rm k,ej} &= 10^{51}\,{\rm erg}, & M_{\rm ej} &= 1\,\rm M_\odot, & v_{\rm ej} &\simeq 10^4~{\rm km\,s^{-1}},
\end{align*}

and representative efficiencies, we consider two cases:  

\begin{enumerate}
    \item High efficiency: $\epsilon_{\rm dep} = 1$, $\epsilon_p = 1$  
    \begin{align}
    M_{\rm strip}^{(E)} &\simeq 4.44\times10^{-1}\,\rm M_\odot,\\
    M_{\rm strip}^{(p)} &\simeq 9.35\times10^{-2}\,\rm M_\odot.
    \end{align}
    
    \item Low efficiency: $\epsilon_{\rm dep} = 0.01$, $\epsilon_p = 0.01$  
    \begin{align}
    M_{\rm strip}^{(E)} &\simeq 4.44\times10^{-3}\,\rm M_\odot,\\
    M_{\rm strip}^{(p)} &\simeq 9.35\times10^{-4}\,\rm M_\odot.
    \end{align}
\end{enumerate}

Both estimates therefore predict $M_{\rm strip} \sim 10^{-3}$--$10^{-1}\,\rm M_\odot$ for these donors. Hence, there is a good possibility that the supernova ejecta could remove most or all of the donor’s He envelope. In that case, the surface would appear predominantly C/O, consistent with the current spectrum of D6-2. Nonetheless, at D6-2’s effective temperature, He may remain spectroscopically hidden, so a residual He layer cannot be completely ruled out. This is consistent with the results of \citet{2019ApJ...887...68B,2024ApJ...973...65W,2025arXiv250812529W}, who also predict that a good fraction of the He envelope can be stripped from the donor following the supernova impact.

\section{Supernova Ia rate and velocity distribution} \label{sec:Population synthesis}

\begin{figure}
    \centering
    \includegraphics[trim = 0 0 0 0 ,width=0.5\textwidth]{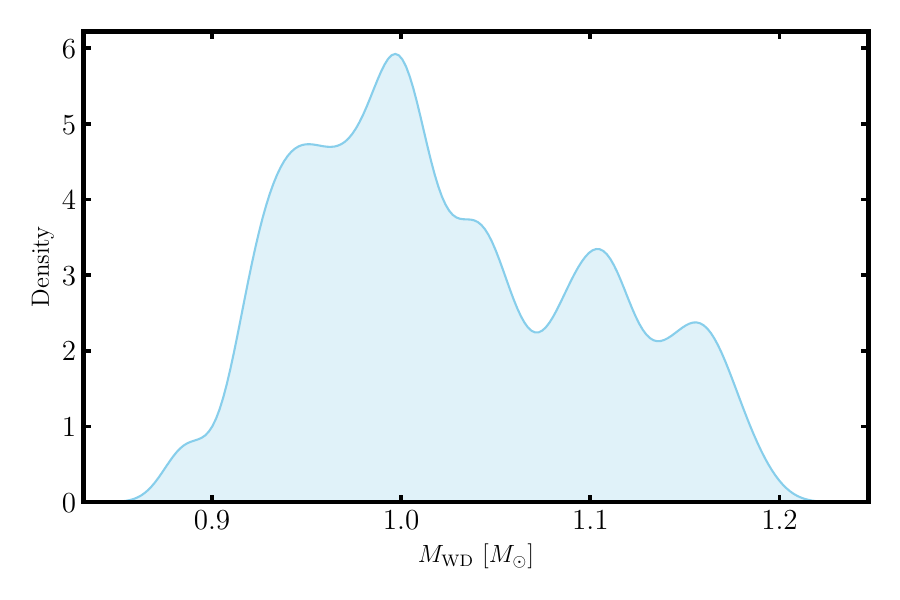}
    \caption{Mass distribution of the white dwarf at the time of the explosion. Most of the white dwarfs undergo double detonation around the peak $\gtrsim 1\rm M_\odot$.}
    \label{fig:ExpWD_mass}
\end{figure}

In this section, we present the estimated Type~Ia supernova rate from the hot subdwarf + white dwarf channel and the corresponding velocity distributions by combining \texttt{MESA} outputs with population synthesis modeling. Since Type~Ia production in these systems requires accurate modeling of the mass-transfer rate and is highly sensitive to the structure of the stars, population synthesis alone cannot capture all relevant physics. Therefore, we take the birth configurations of the binaries from population synthesis, but use detailed \texttt{MESA} calculations to determine which regions of parameter space actually lead to Type~Ia explosions. For this we use the \texttt{MSE}\footnote{\url{ https://github.com/hpreece/mse}.} \citep{2021MNRAS.502.4479H} population synthesis code. The code incorporates detailed prescriptions for stellar evolution, binary interactions (e.g., tides, mass transfer, common-envelope evolution), and dynamical perturbations from higher-order systems and fly-bys. Although \texttt{MSE} can evolve systems of any multiplicity as long as they are initially hierarchical, here we restrict our study to binary systems only. \texttt{MSE} is a publicly available \texttt{C/C++} code with a \texttt{Python} interface. It employs a hybrid scheme that switches between the secular approximation \citep{2016MNRAS.459.2827H,2018MNRAS.476.4139H,2020MNRAS.494.5492H} and direct $N$-body integration \citep{2020MNRAS.492.4131R} when dynamical instability arises. Post-Newtonian (PN) corrections are included up to 2.5\,PN order in the secular regime and up to 3.5\,PN order during direct $N$-body integration. Single-star evolution follows the \texttt{SSE} prescriptions \citep{2000MNRAS.315..543H} based on the evolutionary tracks of \citet{1998MNRAS.298..525P}. Binary interactions are computed using modified \texttt{BSE} algorithms \citep{2002MNRAS.329..897H}. 

\begin{figure}
    \centering
    \includegraphics[trim = 0 0 0 0 ,width=0.5\textwidth]{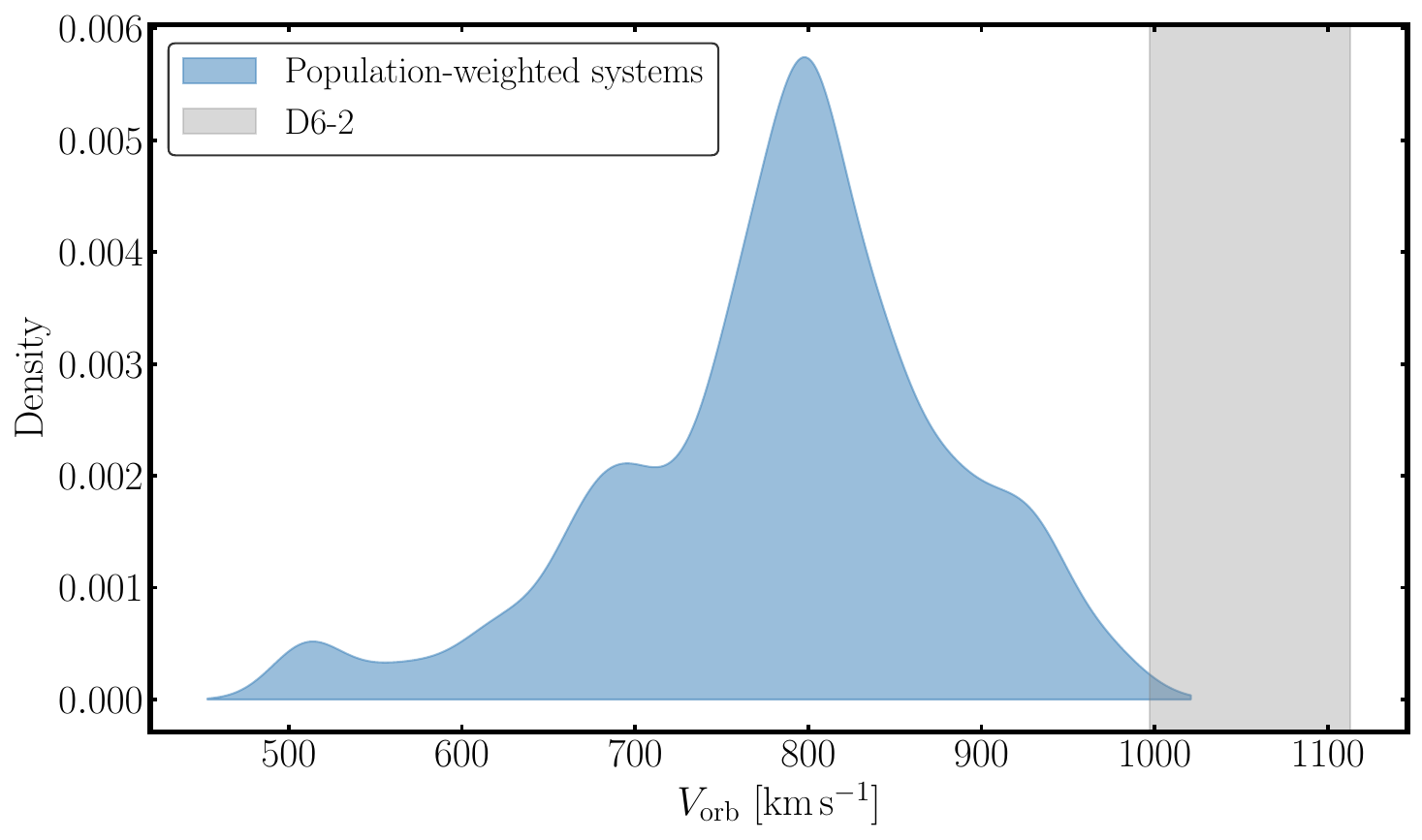}
    \caption{Distribution of ejection velocities for hot subdwarf + white dwarf binaries. The blue curve represents the velocity distribution of the population-weighted distribution after combining MESA and MSE. The gray band marks the observed velocity of D6-2. Both the detailed and population-weighted models predict a substantial fraction of systems with velocities in this range.}
    \label{fig:V_orb}
\end{figure}

For the initial conditions, we generate a synthetic population of binary systems using a Monte Carlo sampling approach. The primary mass ($m_1$) is drawn between $1$ and $10\,\mathrm{M_\odot}$, following the initial mass function of \citet{2001MNRAS.322..231K}, ensuring at least one CO white dwarf forms within a Hubble time. The secondary mass, orbital period, and eccentricity distributions are drawn from the observational fits of \citet{2017ApJS..230...15M}. Systems in which either star initially fills its Roche lobe at periastron are rejected, using the Roche-lobe fit of \citet{1983ApJ...268..368E} and an approximate main-sequence mass–radius relation $R \propto M^{0.7}$.

We evolve the initial binary population using \texttt{MSE} for $14\,\mathrm{Gyr}$. From the resulting sample, we identify systems that form hot subdwarf + white dwarf binaries within our \texttt{MESA} parameter space. The \texttt{MESA} models form a discrete grid in donor mass, accretor mass, and orbital period. However, population-synthesis calculations produce binaries with a distribution of masses and periods that do not exactly fall on this grid. To address this, we use a nearest-neighbor interpolation scheme in all three parameters, weighting deviations in each parameter by the corresponding grid spacing in the \texttt{MESA} models. This allows us to map each population-synthesis binary to the most appropriate \texttt{MESA} evolutionary outcome and determine which systems lead to a white dwarf explosion.

Using the synthetic population and assuming a 50\% binary fraction, we estimate the time-integrated supernova Ia rate from the hot subdwarf white dwarf channel to be ${\sim}(1.69 \pm 0.06)\times10^{-5}\,\rm M_\odot^{-1}$. This corresponds to about 1\% of the time-integrated rate of all observed Type Ia supernovae reported by \citet{2012MNRAS.426.3282M}.

Fig.~\ref{fig:ExpWD_mass} shows the distribution of white dwarf masses at the time of explosion. The initial MESA model grid spans white dwarf masses between 0.7 and 1.1 $\rm M_\odot$. Population synthesis studies predict that the underlying CO white dwarf population peaks near 0.55 $\rm M_\odot$, reflecting the combined effects of the initial mass function and the initial–final mass relation. A secondary peak arises around 0.7–0.8 $\rm M_\odot$ from white dwarfs that have undergone modest pre hot subdwarf accretion in close binaries. During subsequent stable mass transfer from the hot subdwarf companion, these white dwarfs can grow to $\sim1.0,\rm M_\odot$, where double-detonation explosions are most likely to occur. The gradual decline toward higher masses in Fig.~\ref{fig:ExpWD_mass} reflects the decreasing formation probability of initially massive white dwarfs and the limited parameter space available for efficient accretion.

We derive the velocity distribution of the donor stars at the time of explosion, as shown in Fig.~\ref{fig:V_orb}. The blue region denotes the range of orbital velocities spanned by our synthetic population, while the gray shaded region indicates the measured velocity of the observed hypervelocity star D6-2. We can see that the tail of the distribution reaches into the D6-2 range, making it evident that even though rare, there is a possibility to reach this parameter space via hot subdwarf - white dwarf channel. The peak of the velocity distribution is around $\sim 800\,\mathrm{km\,s^{-1}}$.

\section{Discussion}

Our models showed that the hot subdwarf + white dwarf channel can plausibly explain the properties of D6-2. Here, we extend this discussion to consider the implications for the broader population of hypervelocity white dwarfs. We discuss the main uncertainties affecting our binary calculations, compare our proposed He star donor channel for D6-2 with alternative progenitor scenarios, and outline observational prospects to identify additional survivors.

\subsection{Alternative origin scenarios for D6-2 in comparison to hot subdwarf + white dwarf binaries}
\label{subsec:comparison}

Hydrodynamical simulations of supernova impacts provide further insight into extreme outcomes. \cite{2019ApJ...887...68B} used Athena++ hydrodynamical simulations of supernova impact onto semi-degenerate He stars with masses $0.24$ - $0.35\,\rm M_\odot$, showing that although such stars can survive the explosion, their post-impact evolution still tends to produce larger radii and longer contraction times than implied by D6-2. Our models also predict a good fraction of runaway stars with high core He fractions, up to 0.8, which could potentially account for objects like US 708.  \cite{2022A&A...663A..91N} studied the predicted population of runaway He‑sdO/B stars ejected from single‑degenerate He‑donor supernovae and noted that D6‑2’s relatively low velocity is inconsistent with a double‑degenerate ejection scenario, suggesting it could instead be a He star or a proto–white dwarf.

\cite{2025A&A...693A.114B} modeled the cooling of $0.5$--$1.1\,\rm M_\odot$ inflated CO white dwarf donors using MESA with entropy profiles calibrated from AREPO hydrodynamical simulations, demonstrating that these donors contract to $\lesssim 0.01\,\rm R_\odot$ within $\sim 10^4$~yr, which is in tension with the observed inflated state with D6-2. Since their models invoke a D6 scenario, the donor white dwarfs are expected to be ejected with much higher velocities ($>1500\,\rm km\,s^{-1}$), compared to the observed velocity of D6-2. This scenario was further explored by \citet{2025NatAs.tmp..168G}, who studied mergers of hybrid He–CO white dwarfs that produce $\approx 0.5,\mathrm{M_\odot}$ remnants with velocities of order $2{,}000~\mathrm{km\,s^{-1}}$, inconsistent with the observed velocity of D6-2. Similarly, \cite{2025arXiv251011781P} and \cite{2025arXiv251012197B} performed violent merger simulations starting from degenerate CO white dwarfs, resulting in initially non-degenerate hypervelocity remnants that eventually cool into runaway white dwarfs. These pathways, however, typically produce ejection velocities much higher than D6-2 and, like the semi-degenerate He white dwarf models, yield contraction timescales, luminosities, and residual He fractions that remain in tension with the inferred properties of D6-2.  \citet{2025ApJ...982....6S} examined the evolution of hypervelocity survivors from double white dwarf detonations and showed that the cooler D6 stars (including D6-2) could be explained by Kelvin–Helmholtz contraction of He or CO white dwarfs that underwent significant mass loss prior to explosion.  

In particular, \cite{2024ApJ...973...65W} and \citealt{2025arXiv250812529W} focused on semi-degenerate He white dwarfs as potential progenitors for D6-2. While their higher-mass models produce remnants that remain inflated and luminous for longer than D6-2’s observed compact and faint state, the lowest-mass He WD models ($\approx 0.08-0.10,\mathrm{M_\odot}$) are in close agreement with D6-2’s observed characteristics.

If D6-2 originates from a He white dwarf + CO white dwarf channel, it would have to occupy the extreme low-velocity tail of the distribution of surviving donors \citep[e.g.,][]{2018ApJ...865...15S}, implying the existence of more objects with somewhat faster velocities in the range $\sim 1000$–$1500\,\mathrm{km\,s^{-1}}$. Its low velocity is inconsistent with formation from a CO white dwarf + CO white dwarf channel, which typically produces faster ejection speeds.  

In contrast, our hot subdwarf + CO white dwarf scenario predicts essentially the opposite: D6-2 represents the extreme high-velocity tail of a distribution of former subdwarfs that survived the explosion. Depending on the system parameters, these surviving donors may still appear as subdwarfs (e.g., US~708) or have already cooled toward white dwarf-like properties (e.g., D6). The bulk of this predicted population would occupy intermediate velocities, conservatively in the range $\sim 600$–$900\,\mathrm{km\,s^{-1}}$.

Interestingly, searches for hypervelocity stars have identified some objects in this velocity range, notably the LP~40–365 stars. These stars have been interpreted as partially burned deflagration survivors, although their high velocities and, in some cases, chemical abundances challenge this interpretation. In contrast, if they originated from former subdwarfs impacted by a thermonuclear supernova, the velocities would naturally match the observed range. Whether such a scenario can also reproduce the surface abundances of LP~40 stars depends on the degree of accreted supernova ejecta and will need to be tested with future 3D impact simulations.

\subsection{Observing runaway Stars from the Hot Subdwarf + White Dwarf Channel}
\label{subsec:search}

The predicted survivors from the hot subdwarf + white dwarf channel occupy a unique and largely unexplored region of the hypervelocity and runaway white dwarf parameter space. Their detection requires combining precise astrometry, kinematics, and spectroscopy.

Searches for thermonuclear runaway remnants typically begin with astrometric selection from \textit{Gaia} data. Candidates are identified by their large tangential velocities, and relative parallax uncertainties. Cross-matching with photometric catalogues such as SDSS or Pan-STARRS helps to isolate blue, compact objects consistent with hot subdwarfs or white dwarfs. Spectroscopic follow-up is then required to obtain radial velocities, atmospheric parameters, and chemical compositions, enabling full three-dimensional velocity reconstruction in the Galactic rest frame.

A key difficulty lies in distinguishing true supernova remnants from kinematic interlopers. Old halo and thick-disk white dwarfs can reach velocities up to $\sim 400 –500\,\mathrm{km\,s^{-1}}$ \citep[e.g.][]{2023OJAp....6E..28E,2024A&A...690A.368G}, and measurement errors in parallax or radial velocity can inflate inferred tangential speeds, creating artificial hypervelocity outliers. As a result, contamination increases sharply below $\sim$1000~km\,s$^{-1}$, where many bound halo objects reside. Genuine runaway remnants can be confirmed only through accurate 3D orbits showing they are unbound and do not originate from the Galactic Centre, as expected for supernova ejecta.

Spectroscopic diagnostics provide a decisive test. Our models predict compact, C/O dominated transitioning white dwarfs with little or no He, distinguishing them from He rich hot subdwarf stars. However, depending on the explosion geometry and coupling efficiency, part of the supernova ejecta may be deposited onto the donor surface. Such fallback can enrich the outer layers with intermediate-mass elements (e.g. O, Ne, Mg, Si), potentially producing atmospheres resembling LP~40–365 like stars. This suggests that LP~40–365 and D6–2 like objects may represent different observational outcomes of the same hot subdwarf + white dwarf channel, set by the extent of He stripping and ejecta pollution. High-resolution spectra will be essential to test this prediction and constrain the diversity of surface compositions among thermonuclear runaway remnants.

Given that D6–2 is observed at a velocity of $\sim$1000\,\rm km\,s$^{-1}$ and that our population synthesis predicts a peak near $\sim$800\,\rm km\,s$^{-1}$, a comparable number of slightly slower survivors should exist. Because this velocity range overlaps with the most contaminated/faint region of the Galactic velocity distribution, identifying such remnants will require careful statistical vetting and targeted spectroscopic searches in both existing and upcoming large-scale surveys.

\subsection{Model uncertainities}
\label{sec:uncertainities}

All model uncertainties discussed in \citet{2024arXiv241108099R} apply here as well, and we refer the reader to that work for a detailed discussion. In particular, the calculation of orbital velocities is affected by assumptions regarding the conservativeness of mass transfer and the efficiency of tidal interactions. In our binary models, we adopt fully conservative mass transfer with no systemic mass loss, which serves as an initial approximation given that the He retention efficiency of the accretor remains poorly constrained. Future work is needed to quantify the effects of non-conservative mass transfer, including mass and angular-momentum losses due to possible ejection from the system. Accounting for non-conservative mass transfer may facilitate more efficient removal of the donor’s residual He envelope, producing thinner outer layers of the donor stars at the time of explosion of the white dwarf and a composition more consistent with the C/O-rich surfaces observed in D6-2-like objects. We further assume that angular momentum accreted onto the white dwarf is efficiently redistributed back to the orbit through tidal coupling. Because these systems are extremely compact and expected to be tidally synchronized, this approximation is supported by studies such as \citet{2014MNRAS.444.3488F}, which suggest that tidal torques can prevent the accretor from spinning up significantly beyond the orbital frequency.

When normalizing our Type~Ia supernova rates, we adopt the multiplicity fractions from \citet{2017ApJS..230...15M}. Uncertainties in these fractions, as well as in the assumed upper and lower mass limits of the synthesized stellar population, propagate into systematic uncertainties in the overall rate normalization, which we neglect here for simplicity. More broadly, the general uncertainties affecting single and binary stellar evolution models also apply to this work. While the ejection velocities are generally similar to the pre-explosion orbital velocities, small differences may arise from the impact process\citep{2019ApJ...887...68B,2024OJAp....7E...7B,2025A&A...693A.114B,2025arXiv251011781P,2025arXiv251012197B}. We plan to perform detailed 3D hydrodynamical impact simulations to explore the explosion dynamics in greater detail, which will also allow us to determine the ejection velocities more precisely.

\section{Conclusions}
\label{sec:conclusions}

We have investigated the evolution of hot subdwarf + CO white dwarf binaries as potential progenitors of thermonuclear supernovae and their surviving runaway companions. Using detailed binary evolution models computed with \textsc{MESA}, combined with population synthesis calculations from \textsc{MSE}, we explored the conditions under which the donor evolves into a compact, nearly degenerate object prior to the explosion of the accretor. In all cases, the explosions correspond to double detonations of sub-Chandrasekhar-mass CO white dwarfs. Our main findings are as follows:

\begin{enumerate}
 \item The single-degenerate channel does not exclusively produce non-degenerate survivors. In many hot subdwarf + white dwarf systems, the donor evolves into a transitioning white dwarf before the detonation of the accretor. These donors are compact and nearly degenerate at the time of explosion.

\item The ejection velocities of these degenerate donors naturally span an intermediate range of $\sim 450$–$1000\,\mathrm{km\,s^{-1}}$, with the upper end matching the observed velocity of D6-2. This range bridges the gap between He star and double white dwarf progenitor systems

\item The effective temperature of D6-2 is sufficiently low that any residual He would not produce detectable spectral lines. Even if a thin He envelope remains post-explosion, analytical estimates indicate that it can be potentially stripped by the supernova ejecta, resulting in a surface composition dominated by carbon and oxygen, consistent with observations.

\item This study estimates a Type~Ia supernova rate of ${\sim}(1.69 \pm 0.06)\times10^{-5}\,\rm M_\odot^{-1}$, indicating that hot subdwarf + white dwarf binaries form at a rate sufficient to contribute a small but non-negligible fraction of the total Type~Ia population, and potentially account for a good fraction of the observed hypervelocity white dwarfs.

\item The orbital velocities of donor stars in similar systems are consistent with the inferred ejection velocities of LP 40–365 stars. Whether this channel also reproduces their surface abundances will be tested in our future 3D supernova impact simulations.

\end{enumerate}

These results identify the hot subdwarf + white dwarf channel as a viable evolutionary bridge between He star and double white dwarf progenitor systems. Future work will focus on hydrodynamical simulations of the supernova impact and post-explosion evolution of the surviving donor to predict surface properties such as composition, luminosity, and temperature. Discovery of additional intermediate-velocity, CO-rich runaway white dwarfs would provide strong empirical support for this evolutionary pathway.

\begin{acknowledgements}
ASR thanks Evan Bauer and Selma de Mink for helpful comments. We also express our gratitude to the Kavli Foundation for funding the MPA/Kavli Summer Program 2023, during which this collaboration became possible. A.B. was supported by the Deutsche Forschungsgemeinschaft (DFG) through grant GE2506/18-1. 
\end{acknowledgements}

\clearpage

\bibliographystyle{aa} 
\bibliography{paper} 

@ARTICLE{1969ApJ...157..623C,
       author = {{Colgate}, Stirling A. and {McKee}, Chester},
        title = "{Early Supernova Luminosity}",
      journal = {\apj},
         year = 1969,
        month = aug,
       volume = {157},
        pages = {623},
          doi = {10.1086/150102},
       adsurl = {https://ui.adsabs.harvard.edu/abs/1969ApJ...157..623C}
}

@ARTICLE{1982ApJ...253..785A,
       author = {{Arnett}, W.~D.},
        title = "{Type I supernovae. I - Analytic solutions for the early part of the light curve}",
      journal = {\apj},
         year = 1982,
        month = feb,
       volume = {253},
        pages = {785-797},
          doi = {10.1086/159681},
       adsurl = {https://ui.adsabs.harvard.edu/abs/1982ApJ...253..785A}
}

@ARTICLE{2013FrPhy...8..116H,
       author = {{Hillebrandt}, W. and {Kromer}, M. and {R{\"o}pke}, F.~K. and {Ruiter}, A.~J.},
        title = "{Towards an understanding of Type Ia supernovae from a synthesis of theory and observations}",
      journal = {Frontiers of Physics},
         year = 2013,
        month = apr,
       volume = {8},
       number = {2},
        pages = {116-143},
          doi = {10.1007/s11467-013-0303-2},
archivePrefix = {arXiv},
       eprint = {1302.6420},
 primaryClass = {astro-ph.CO},
       adsurl = {https://ui.adsabs.harvard.edu/abs/2013FrPhy...8..116H}
}

@ARTICLE{1998AJ....116.1009R,
       author = {{Riess}, Adam G. and {Filippenko}, Alexei V. and {Challis}, Peter and {Clocchiatti}, Alejandro and {Diercks}, Alan and {Garnavich}, Peter M. and {Gilliland}, Ron L. and {Hogan}, Craig J. and {Jha}, Saurabh and {Kirshner}, Robert P. and {Leibundgut}, B. and {Phillips}, M.~M. and {Reiss}, David and {Schmidt}, Brian P. and {Schommer}, Robert A. and {Smith}, R. Chris and {Spyromilio}, J. and {Stubbs}, Christopher and {Suntzeff}, Nicholas B. and {Tonry}, John},
        title = "{Observational Evidence from Supernovae for an Accelerating Universe and a Cosmological Constant}",
      journal = {\aj},
         year = 1998,
        month = sep,
       volume = {116},
       number = {3},
        pages = {1009-1038},
          doi = {10.1086/300499},
archivePrefix = {arXiv},
       eprint = {astro-ph/9805201},
 primaryClass = {astro-ph},
       adsurl = {https://ui.adsabs.harvard.edu/abs/1998AJ....116.1009R}
}

@ARTICLE{1999ApJ...517..565P,
       author = {{Perlmutter}, S. and {Aldering}, G. and {Goldhaber}, G. and {Knop}, R.~A. and {Nugent}, P. and {Castro}, P.~G. and {Deustua}, S. and {Fabbro}, S. and {Goobar}, A. and {Groom}, D.~E. and {Hook}, I.~M. and {Kim}, A.~G. and {Kim}, M.~Y. and {Lee}, J.~C. and {Nunes}, N.~J. and {Pain}, R. and {Pennypacker}, C.~R. and {Quimby}, R. and {Lidman}, C. and {Ellis}, R.~S. and {Irwin}, M. and {McMahon}, R.~G. and {Ruiz-Lapuente}, P. and {Walton}, N. and {Schaefer}, B. and {Boyle}, B.~J. and {Filippenko}, A.~V. and {Matheson}, T. and {Fruchter}, A.~S. and {Panagia}, N. and {Newberg}, H.~J.~M. and {Couch}, W.~J. and {Project}, The Supernova Cosmology},
        title = "{Measurements of {\ensuremath{\Omega}} and {\ensuremath{\Lambda}} from 42 High-Redshift Supernovae}",
      journal = {\apj},
         year = 1999,
        month = jun,
       volume = {517},
       number = {2},
        pages = {565-586},
          doi = {10.1086/307221},
archivePrefix = {arXiv},
       eprint = {astro-ph/9812133},
 primaryClass = {astro-ph},
       adsurl = {https://ui.adsabs.harvard.edu/abs/1999ApJ...517..565P}
}

@ARTICLE{1998ApJ...507...46S,
       author = {{Schmidt}, Brian P. and {Suntzeff}, Nicholas B. and {Phillips}, M.~M. and {Schommer}, Robert A. and {Clocchiatti}, Alejandro and {Kirshner}, Robert P. and {Garnavich}, Peter and {Challis}, Peter and {Leibundgut}, B. and {Spyromilio}, J. and {Riess}, Adam G. and {Filippenko}, Alexei V. and {Hamuy}, Mario and {Smith}, R. Chris and {Hogan}, Craig and {Stubbs}, Christopher and {Diercks}, Alan and {Reiss}, David and {Gilliland}, Ron and {Tonry}, John and {Maza}, Jos{\'e} and {Dressler}, A. and {Walsh}, J. and {Ciardullo}, R.},
        title = "{The High-Z Supernova Search: Measuring Cosmic Deceleration and Global Curvature of the Universe Using Type IA Supernovae}",
      journal = {\apj},
         year = 1998,
        month = nov,
       volume = {507},
       number = {1},
        pages = {46-63},
          doi = {10.1086/306308},
archivePrefix = {arXiv},
       eprint = {astro-ph/9805200},
 primaryClass = {astro-ph},
       adsurl = {https://ui.adsabs.harvard.edu/abs/1998ApJ...507...46S}
}

@ARTICLE{2023RAA....23h2001L,
       author = {{Liu}, Zheng-Wei and {R{\"o}pke}, Friedrich K. and {Han}, Zhanwen},
        title = "{Type Ia Supernova Explosions in Binary Systems: A Review}",
      journal = {Research in Astronomy and Astrophysics},
         year = 2023,
        month = aug,
       volume = {23},
       number = {8},
          eid = {082001},
        pages = {082001},
          doi = {10.1088/1674-4527/acd89e},
archivePrefix = {arXiv},
       eprint = {2305.13305},
 primaryClass = {astro-ph.HE},
       adsurl = {https://ui.adsabs.harvard.edu/abs/2023RAA....23h2001L}
}

@ARTICLE{2025A&ARv..33....1R,
       author = {{Ruiter}, Ashley Jade and {Seitenzahl}, Ivo Rolf},
        title = "{Type Ia supernova progenitors: a contemporary view of a long-standing puzzle}",
      journal = {\aapr},
         year = 2025,
        month = dec,
       volume = {33},
       number = {1},
          eid = {1},
        pages = {1},
          doi = {10.1007/s00159-024-00158-9},
archivePrefix = {arXiv},
       eprint = {2412.01766},
 primaryClass = {astro-ph.SR},
       adsurl = {https://ui.adsabs.harvard.edu/abs/2025A&ARv..33....1R}
}

@ARTICLE{1973ApJ...186.1007W,
       author = {{Whelan}, John and {Iben}, Jr., Icko},
        title = "{Binaries and Supernovae of Type I}",
      journal = {\apj},
         year = 1973,
        month = dec,
       volume = {186},
        pages = {1007-1014},
          doi = {10.1086/152565},
       adsurl = {https://ui.adsabs.harvard.edu/abs/1973ApJ...186.1007W}
}

@ARTICLE{1982ApJ...257..780N,
       author = {{Nomoto}, K.},
        title = "{Accreting white dwarf models for type I supernovae. II. Off-center detonation supernovae.}",
      journal = {\apj},
         year = 1982,
        month = jun,
       volume = {257},
        pages = {780-792},
          doi = {10.1086/160031},
       adsurl = {https://ui.adsabs.harvard.edu/abs/1982ApJ...257..780N}
}

@ARTICLE{1982ApJ...253..798N,
       author = {{Nomoto}, K.},
        title = "{Accreting white dwarf models for type I supernovae. I - Presupernova evolution and triggering mechanisms}",
      journal = {\apj},
         year = 1982,
        month = feb,
       volume = {253},
        pages = {798-810},
          doi = {10.1086/159682},
       adsurl = {https://ui.adsabs.harvard.edu/abs/1982ApJ...253..798N}
}

@ARTICLE{1984ApJ...286..644N,
       author = {{Nomoto}, K. and {Thielemann}, F. -K. and {Yokoi}, K.},
        title = "{Accreting white dwarf models for type I supernovae. III. Carbon deflagration supernovae.}",
      journal = {\apj},
         year = 1984,
        month = nov,
       volume = {286},
        pages = {644-658},
          doi = {10.1086/162639},
       adsurl = {https://ui.adsabs.harvard.edu/abs/1984ApJ...286..644N}
}

@ARTICLE{1984ApJ...277..355W,
       author = {{Webbink}, R.~F.},
        title = "{Double white dwarfs as progenitors of R Coronae Borealis stars and type I supernovae.}",
      journal = {\apj},
         year = 1984,
        month = feb,
       volume = {277},
        pages = {355-360},
          doi = {10.1086/161701},
       adsurl = {https://ui.adsabs.harvard.edu/abs/1984ApJ...277..355W}
}

@ARTICLE{1984ApJ...284..719I,
       author = {{Iben}, Jr., I. and {Tutukov}, A.~V.},
        title = "{The evolution of low-mass close binaries influenced by the radiation of gravitational waves and by a magnetic stellar wind}",
      journal = {\apj},
         year = 1984,
        month = sep,
       volume = {284},
        pages = {719-744},
          doi = {10.1086/162455},
       adsurl = {https://ui.adsabs.harvard.edu/abs/1984ApJ...284..719I}
}

@ARTICLE{2009ApJ...701.1665K,
       author = {{Kerzendorf}, Wolfgang E. and {Schmidt}, Brian P. and {Asplund}, M. and {Nomoto}, Ken'ichi and {Podsiadlowski}, Ph. and {Frebel}, Anna and {Fesen}, Robert A. and {Yong}, David},
        title = "{Subaru High-Resolution Spectroscopy of Star G in the Tycho Supernova Remnant}",
      journal = {\apj},
         year = 2009,
        month = aug,
       volume = {701},
       number = {2},
        pages = {1665-1672},
          doi = {10.1088/0004-637X/701/2/1665},
archivePrefix = {arXiv},
       eprint = {0906.0982},
 primaryClass = {astro-ph.SR},
       adsurl = {https://ui.adsabs.harvard.edu/abs/2009ApJ...701.1665K}
}

@ARTICLE{2012Natur.481..164S,
       author = {{Schaefer}, Bradley E. and {Pagnotta}, Ashley},
        title = "{An absence of ex-companion stars in the type Ia supernova remnant SNR 0509-67.5}",
      journal = {\nat},
         year = 2012,
        month = jan,
       volume = {481},
       number = {7380},
        pages = {164-166},
          doi = {10.1038/nature10692},
       adsurl = {https://ui.adsabs.harvard.edu/abs/2012Natur.481..164S}
}

@ARTICLE{2018ApJ...862..124R,
       author = {{Ruiz-Lapuente}, Pilar and {Damiani}, Francesco and {Bedin}, Luigi and {Gonz{\'a}lez Hern{\'a}ndez}, Jonay I. and {Galbany}, Llu{\'\i}s and {Pritchard}, John and {Canal}, Ramon and {M{\'e}ndez}, Javier},
        title = "{No Surviving Companion in Kepler's Supernova}",
      journal = {\apj},
         year = 2018,
        month = aug,
       volume = {862},
       number = {2},
          eid = {124},
        pages = {124},
          doi = {10.3847/1538-4357/aac9c4},
archivePrefix = {arXiv},
       eprint = {1711.00876},
 primaryClass = {astro-ph.SR},
       adsurl = {https://ui.adsabs.harvard.edu/abs/2018ApJ...862..124R}
}

@ARTICLE{2005A&A...444L..61H,
       author = {{Hirsch}, H.~A. and {Heber}, U. and {O'Toole}, S.~J. and {Bresolin}, F.},
        title = "{<ASTROBJ>US 708</ASTROBJ> - an unbound hyper-velocity subluminous O star}",
      journal = {\aap},
         year = 2005,
        month = dec,
       volume = {444},
       number = {3},
        pages = {L61-L64},
          doi = {10.1051/0004-6361:200500212},
archivePrefix = {arXiv},
       eprint = {astro-ph/0511323},
 primaryClass = {astro-ph},
       adsurl = {https://ui.adsabs.harvard.edu/abs/2005A&A...444L..61H}
}

@ARTICLE{2015Sci...347.1126G,
       author = {{Geier}, S. and {F{\"u}rst}, F. and {Ziegerer}, E. and {Kupfer}, T. and {Heber}, U. and {Irrgang}, A. and {Wang}, B. and {Liu}, Z. and {Han}, Z. and {Sesar}, B. and {Levitan}, D. and {Kotak}, R. and {Magnier}, E. and {Smith}, K. and {Burgett}, W.~S. and {Chambers}, K. and {Flewelling}, H. and {Kaiser}, N. and {Wainscoat}, R. and {Waters}, C.},
        title = "{The fastest unbound star in our Galaxy ejected by a thermonuclear supernova}",
      journal = {Science},
         year = 2015,
        month = mar,
       volume = {347},
       number = {6226},
        pages = {1126-1128},
          doi = {10.1126/science.1259063},
archivePrefix = {arXiv},
       eprint = {1503.01650},
 primaryClass = {astro-ph.SR},
       adsurl = {https://ui.adsabs.harvard.edu/abs/2015Sci...347.1126G}
}

@ARTICLE{2015ApJ...804...49B,
       author = {{Brown}, Warren R. and {Anderson}, Jay and {Gnedin}, Oleg Y. and {Bond}, Howard E. and {Geller}, Margaret J. and {Kenyon}, Scott J.},
        title = "{Proper Motions and Trajectories for 16 Extreme Runaway and Hypervelocity Stars}",
      journal = {\apj},
         year = 2015,
        month = may,
       volume = {804},
       number = {1},
          eid = {49},
        pages = {49},
          doi = {10.1088/0004-637X/804/1/49},
archivePrefix = {arXiv},
       eprint = {1502.05069},
 primaryClass = {astro-ph.SR},
       adsurl = {https://ui.adsabs.harvard.edu/abs/2015ApJ...804...49B}
}

@ARTICLE{2018ApJ...865...15S,
       author = {{Shen}, Ken J. and {Boubert}, Douglas and {G{\"a}nsicke}, Boris T. and {Jha}, Saurabh W. and {Andrews}, Jennifer E. and {Chomiuk}, Laura and {Foley}, Ryan J. and {Fraser}, Morgan and {Gromadzki}, Mariusz and {Guillochon}, James and {Kotze}, Marissa M. and {Maguire}, Kate and {Siebert}, Matthew R. and {Smith}, Nathan and {Strader}, Jay and {Badenes}, Carles and {Kerzendorf}, Wolfgang E. and {Koester}, Detlev and {Kromer}, Markus and {Miles}, Broxton and {Pakmor}, R{\"u}diger and {Schwab}, Josiah and {Toloza}, Odette and {Toonen}, Silvia and {Townsley}, Dean M. and {Williams}, Brian J.},
        title = "{Three Hypervelocity White Dwarfs in Gaia DR2: Evidence for Dynamically Driven Double-degenerate Double-detonation Type Ia Supernovae}",
      journal = {\apj},
         year = 2018,
        month = sep,
       volume = {865},
       number = {1},
          eid = {15},
        pages = {15},
          doi = {10.3847/1538-4357/aad55b},
archivePrefix = {arXiv},
       eprint = {1804.11163},
 primaryClass = {astro-ph.SR},
       adsurl = {https://ui.adsabs.harvard.edu/abs/2018ApJ...865...15S}
}

@ARTICLE{2023OJAp....6E..28E,
       author = {{El-Badry}, Kareem and {Shen}, Ken J. and {Chandra}, Vedant and {Bauer}, Evan B. and {Fuller}, Jim and {Strader}, Jay and {Chomiuk}, Laura and {Naidu}, Rohan P. and {Caiazzo}, Ilaria and {Rodriguez}, Antonio C. and {Nagarajan}, Pranav and {Yamaguchi}, Natsuko and {Vanderbosch}, Zachary P. and {Roulston}, Benjamin R. and {G{\"a}nsicke}, Boris and {Han}, Jiwon Jesse and {Burdge}, Kevin B. and {Filippenko}, Alexei V. and {Brink}, Thomas G. and {Zheng}, WeiKang},
        title = "{The fastest stars in the Galaxy}",
      journal = {The Open Journal of Astrophysics},
         year = 2023,
        month = jul,
       volume = {6},
          eid = {28},
        pages = {28},
          doi = {10.21105/astro.2306.03914},
archivePrefix = {arXiv},
       eprint = {2306.03914},
 primaryClass = {astro-ph.SR},
       adsurl = {https://ui.adsabs.harvard.edu/abs/2023OJAp....6E..28E}
}

@ARTICLE{2008ApJ...677L.109H,
       author = {{Han}, Z.},
        title = "{Companion Stars of Type Ia Supernovae}",
      journal = {\apjl},
         year = 2008,
        month = apr,
       volume = {677},
       number = {2},
        pages = {L109},
          doi = {10.1086/588191},
archivePrefix = {arXiv},
       eprint = {0803.1986},
 primaryClass = {astro-ph},
       adsurl = {https://ui.adsabs.harvard.edu/abs/2008ApJ...677L.109H}
}

@ARTICLE{2024arXiv241108099R,
       author = {{Rajamuthukumar}, Abinaya Swaruba and {Bauer}, Evan B. and {Justham}, Stephen and {Pakmor}, R{\"u}diger and {de Mink}, Selma E. and {Neunteufel}, Patrick},
        title = "{Evolution of binaries containing a hot subdwarf and a white dwarf to double white dwarfs, and double detonation supernovae with hypervelocity runaway stars}",
      journal = {arXiv e-prints},
         year = 2024,
        month = nov,
          eid = {arXiv:2411.08099},
        pages = {arXiv:2411.08099},
          doi = {10.48550/arXiv.2411.08099},
archivePrefix = {arXiv},
       eprint = {2411.08099},
 primaryClass = {astro-ph.SR},
       adsurl = {https://ui.adsabs.harvard.edu/abs/2024arXiv241108099R}
}

@ARTICLE{2025NatAs.tmp..168G,
       author = {{Glanz}, Hila and {Perets}, Hagai B. and {Bhat}, Aakash and {Pakmor}, Ruediger},
        title = "{The origin of hypervelocity white dwarfs in the merger disruption of He-C-O white dwarfs}",
      journal = {Nature Astronomy},
         year = 2025,
        month = aug,
          doi = {10.1038/s41550-025-02633-4},
archivePrefix = {arXiv},
       eprint = {2410.17306},
 primaryClass = {astro-ph.HE},
       adsurl = {https://ui.adsabs.harvard.edu/abs/2025NatAs.tmp..168G}
}

@ARTICLE{2024ApJ...973...65W,
       author = {{Wong}, Tin Long Sunny and {White}, Christopher J. and {Bildsten}, Lars},
        title = "{Shocking and Mass Loss of Compact Donor Stars in Type Ia Supernovae}",
      journal = {\apj},
         year = 2024,
        month = sep,
       volume = {973},
       number = {1},
          eid = {65},
        pages = {65},
          doi = {10.3847/1538-4357/ad6a11},
archivePrefix = {arXiv},
       eprint = {2408.00125},
 primaryClass = {astro-ph.SR},
       adsurl = {https://ui.adsabs.harvard.edu/abs/2024ApJ...973...65W}
}

@ARTICLE{2025arXiv250812529W,
       author = {{Wong}, Tin Long Sunny and {Bildsten}, Lars},
        title = "{Mass Loss and Subsequent Thermal Evolution of Surviving Helium White Dwarfs Shocked by Thermonuclear Supernovae}",
      journal = {arXiv e-prints},
         year = 2025,
        month = aug,
          eid = {arXiv:2508.12529},
        pages = {arXiv:2508.12529},
          doi = {10.48550/arXiv.2508.12529},
archivePrefix = {arXiv},
       eprint = {2508.12529},
 primaryClass = {astro-ph.SR},
       adsurl = {https://ui.adsabs.harvard.edu/abs/2025arXiv250812529W}
}

@ARTICLE{2011ApJS..192....3P,
       author = {{Paxton}, Bill and {Bildsten}, Lars and {Dotter}, Aaron and {Herwig}, Falk and {Lesaffre}, Pierre and {Timmes}, Frank},
        title = "{Modules for Experiments in Stellar Astrophysics (MESA)}",
      journal = {\apjs},
         year = 2011,
        month = jan,
       volume = {192},
       number = {1},
          eid = {3},
        pages = {3},
          doi = {10.1088/0067-0049/192/1/3},
archivePrefix = {arXiv},
       eprint = {1009.1622},
 primaryClass = {astro-ph.SR},
       adsurl = {https://ui.adsabs.harvard.edu/abs/2011ApJS..192....3P}
}

@ARTICLE{2013ApJS..208....4P,
       author = {{Paxton}, Bill and {Cantiello}, Matteo and {Arras}, Phil and {Bildsten}, Lars and {Brown}, Edward F. and {Dotter}, Aaron and {Mankovich}, Christopher and {Montgomery}, M.~H. and {Stello}, Dennis and {Timmes}, F.~X. and {Townsend}, Richard},
        title = "{Modules for Experiments in Stellar Astrophysics (MESA): Planets, Oscillations, Rotation, and Massive Stars}",
      journal = {\apjs},
         year = 2013,
        month = sep,
       volume = {208},
       number = {1},
          eid = {4},
        pages = {4},
          doi = {10.1088/0067-0049/208/1/4},
archivePrefix = {arXiv},
       eprint = {1301.0319},
 primaryClass = {astro-ph.SR},
       adsurl = {https://ui.adsabs.harvard.edu/abs/2013ApJS..208....4P}
}

@ARTICLE{2015ApJS..220...15P,
       author = {{Paxton}, Bill and {Marchant}, Pablo and {Schwab}, Josiah and {Bauer}, Evan B. and {Bildsten}, Lars and {Cantiello}, Matteo and {Dessart}, Luc and {Farmer}, R. and {Hu}, H. and {Langer}, N. and {Townsend}, R.~H.~D. and {Townsley}, Dean M. and {Timmes}, F.~X.},
        title = "{Modules for Experiments in Stellar Astrophysics (MESA): Binaries, Pulsations, and Explosions}",
      journal = {\apjs},
         year = 2015,
        month = sep,
       volume = {220},
       number = {1},
          eid = {15},
        pages = {15},
          doi = {10.1088/0067-0049/220/1/15},
archivePrefix = {arXiv},
       eprint = {1506.03146},
 primaryClass = {astro-ph.SR},
       adsurl = {https://ui.adsabs.harvard.edu/abs/2015ApJS..220...15P}
}

@ARTICLE{2018ApJS..234...34P,
       author = {{Paxton}, Bill and {Schwab}, Josiah and {Bauer}, Evan B. and {Bildsten}, Lars and {Blinnikov}, Sergei and {Duffell}, Paul and {Farmer}, R. and {Goldberg}, Jared A. and {Marchant}, Pablo and {Sorokina}, Elena and {Thoul}, Anne and {Townsend}, Richard H.~D. and {Timmes}, F.~X.},
        title = "{Modules for Experiments in Stellar Astrophysics (MESA): Convective Boundaries, Element Diffusion, and Massive Star Explosions}",
      journal = {\apjs},
         year = 2018,
        month = feb,
       volume = {234},
       number = {2},
          eid = {34},
        pages = {34},
          doi = {10.3847/1538-4365/aaa5a8},
archivePrefix = {arXiv},
       eprint = {1710.08424},
 primaryClass = {astro-ph.SR},
       adsurl = {https://ui.adsabs.harvard.edu/abs/2018ApJS..234...34P}
}

@ARTICLE{2019ApJS..243...10P,
       author = {{Paxton}, Bill and {Smolec}, R. and {Schwab}, Josiah and {Gautschy}, A. and {Bildsten}, Lars and {Cantiello}, Matteo and {Dotter}, Aaron and {Farmer}, R. and {Goldberg}, Jared A. and {Jermyn}, Adam S. and {Kanbur}, S.~M. and {Marchant}, Pablo and {Thoul}, Anne and {Townsend}, Richard H.~D. and {Wolf}, William M. and {Zhang}, Michael and {Timmes}, F.~X.},
        title = "{Modules for Experiments in Stellar Astrophysics (MESA): Pulsating Variable Stars, Rotation, Convective Boundaries, and Energy Conservation}",
      journal = {\apjs},
         year = 2019,
        month = jul,
       volume = {243},
       number = {1},
          eid = {10},
        pages = {10},
          doi = {10.3847/1538-4365/ab2241},
archivePrefix = {arXiv},
       eprint = {1903.01426},
 primaryClass = {astro-ph.SR},
       adsurl = {https://ui.adsabs.harvard.edu/abs/2019ApJS..243...10P}
}

@ARTICLE{2023ApJS..265...15J,
       author = {{Jermyn}, Adam S. and {Bauer}, Evan B. and {Schwab}, Josiah and {Farmer}, R. and {Ball}, Warrick H. and {Bellinger}, Earl P. and {Dotter}, Aaron and {Joyce}, Meridith and {Marchant}, Pablo and {Mombarg}, Joey S.~G. and {Wolf}, William M. and {Sunny Wong}, Tin Long and {Cinquegrana}, Giulia C. and {Farrell}, Eoin and {Smolec}, R. and {Thoul}, Anne and {Cantiello}, Matteo and {Herwig}, Falk and {Toloza}, Odette and {Bildsten}, Lars and {Townsend}, Richard H.~D. and {Timmes}, F.~X.},
        title = "{Modules for Experiments in Stellar Astrophysics (MESA): Time-dependent Convection, Energy Conservation, Automatic Differentiation, and Infrastructure}",
      journal = {\apjs},
         year = 2023,
        month = mar,
       volume = {265},
       number = {1},
          eid = {15},
        pages = {15},
          doi = {10.3847/1538-4365/acae8d},
archivePrefix = {arXiv},
       eprint = {2208.03651},
 primaryClass = {astro-ph.SR},
       adsurl = {https://ui.adsabs.harvard.edu/abs/2023ApJS..265...15J}
}

@ARTICLE{1990A&A...236..385K,
       author = {{Kolb}, U. and {Ritter}, H.},
        title = "{A comparative study of the evolution of a close binary using a standard and an improved technique for computing mass transfer.}",
      journal = {\aap},
         year = 1990,
        month = sep,
       volume = {236},
        pages = {385-392},
       adsurl = {https://ui.adsabs.harvard.edu/abs/1990A&A...236..385K}
}

@ARTICLE{1994ApJ...423..371W,
       author = {{Woosley}, S.~E. and {Weaver}, Thomas A.},
        title = "{Sub--Chandrasekhar Mass Models for Type IA Supernovae}",
      journal = {\apj},
         year = 1994,
        month = mar,
       volume = {423},
        pages = {371},
          doi = {10.1086/173813},
       adsurl = {https://ui.adsabs.harvard.edu/abs/1994ApJ...423..371W}
}

@ARTICLE{2017ApJ...845...97B,
       author = {{Bauer}, Evan B. and {Schwab}, Josiah and {Bildsten}, Lars},
        title = "{Electron Captures on $^{14}$N as a Trigger for Helium Shell Detonations}",
      journal = {\apj},
         year = 2017,
        month = aug,
       volume = {845},
       number = {2},
          eid = {97},
        pages = {97},
          doi = {10.3847/1538-4357/aa7ffa},
archivePrefix = {arXiv},
       eprint = {1707.05394},
 primaryClass = {astro-ph.SR},
       adsurl = {https://ui.adsabs.harvard.edu/abs/2017ApJ...845...97B}
}

@ARTICLE{2021MNRAS.502.4479H,
       author = {{Hamers}, Adrian S. and {Rantala}, Antti and {Neunteufel}, Patrick and {Preece}, Holly and {Vynatheya}, Pavan},
        title = "{Multiple Stellar Evolution: a population synthesis algorithm to model the stellar, binary, and dynamical evolution of multiple-star systems}",
      journal = {\mnras},
         year = 2021,
        month = apr,
       volume = {502},
       number = {3},
        pages = {4479-4512},
          doi = {10.1093/mnras/stab287},
archivePrefix = {arXiv},
       eprint = {2011.04513},
 primaryClass = {astro-ph.SR},
       adsurl = {https://ui.adsabs.harvard.edu/abs/2021MNRAS.502.4479H}
}

@ARTICLE{2016MNRAS.459.2827H,
       author = {{Hamers}, Adrian S. and {Portegies Zwart}, Simon F.},
        title = "{Secular dynamics of hierarchical multiple systems composed of nested binaries, with an arbitrary number of bodies and arbitrary hierarchical structure. First applications to multiplanet and multistar systems}",
      journal = {\mnras},
         year = 2016,
        month = jul,
       volume = {459},
       number = {3},
        pages = {2827-2874},
          doi = {10.1093/mnras/stw784},
archivePrefix = {arXiv},
       eprint = {1511.00944},
 primaryClass = {astro-ph.SR},
       adsurl = {https://ui.adsabs.harvard.edu/abs/2016MNRAS.459.2827H}
}

@ARTICLE{2018MNRAS.476.4139H,
       author = {{Hamers}, Adrian S.},
        title = "{Secular dynamics of hierarchical multiple systems composed of nested binaries, with an arbitrary number of bodies and arbitrary hierarchical structure - II. External perturbations: flybys and supernovae}",
      journal = {\mnras},
         year = 2018,
        month = may,
       volume = {476},
       number = {3},
        pages = {4139-4161},
          doi = {10.1093/mnras/sty428},
archivePrefix = {arXiv},
       eprint = {1802.05716},
 primaryClass = {astro-ph.SR},
       adsurl = {https://ui.adsabs.harvard.edu/abs/2018MNRAS.476.4139H}
}

@ARTICLE{2020MNRAS.494.5492H,
       author = {{Hamers}, Adrian S.},
        title = "{Secular dynamics of hierarchical multiple systems composed of nested binaries, with an arbitrary number of bodies and arbitrary hierarchical structure - III. Suborbital effects: hybrid integration techniques and orbit-averaging corrections}",
      journal = {\mnras},
         year = 2020,
        month = jun,
       volume = {494},
       number = {4},
        pages = {5492-5506},
          doi = {10.1093/mnras/staa1084},
archivePrefix = {arXiv},
       eprint = {2004.08327},
 primaryClass = {astro-ph.EP},
       adsurl = {https://ui.adsabs.harvard.edu/abs/2020MNRAS.494.5492H}
}

@ARTICLE{2020MNRAS.492.4131R,
       author = {{Rantala}, Antti and {Pihajoki}, Pauli and {Mannerkoski}, Matias and {Johansson}, Peter H. and {Naab}, Thorsten},
        title = "{MSTAR - a fast parallelized algorithmically regularized integrator with minimum spanning tree coordinates}",
      journal = {\mnras},
         year = 2020,
        month = mar,
       volume = {492},
       number = {3},
        pages = {4131-4148},
          doi = {10.1093/mnras/staa084},
archivePrefix = {arXiv},
       eprint = {2001.03180},
 primaryClass = {astro-ph.IM},
       adsurl = {https://ui.adsabs.harvard.edu/abs/2020MNRAS.492.4131R}
}

@ARTICLE{2000MNRAS.315..543H,
       author = {{Hurley}, Jarrod R. and {Pols}, Onno R. and {Tout}, Christopher A.},
        title = "{Comprehensive analytic formulae for stellar evolution as a function of mass and metallicity}",
      journal = {\mnras},
         year = 2000,
        month = jul,
       volume = {315},
       number = {3},
        pages = {543-569},
          doi = {10.1046/j.1365-8711.2000.03426.x},
archivePrefix = {arXiv},
       eprint = {astro-ph/0001295},
 primaryClass = {astro-ph},
       adsurl = {https://ui.adsabs.harvard.edu/abs/2000MNRAS.315..543H}
}

@ARTICLE{1998MNRAS.298..525P,
       author = {{Pols}, Onno R. and {Schr{\"o}der}, Klaus-Peter and {Hurley}, Jarrod R. and {Tout}, Christopher A. and {Eggleton}, Peter P.},
        title = "{Stellar evolution models for Z = 0.0001 to 0.03}",
      journal = {\mnras},
         year = 1998,
        month = aug,
       volume = {298},
       number = {2},
        pages = {525-536},
          doi = {10.1046/j.1365-8711.1998.01658.x},
       adsurl = {https://ui.adsabs.harvard.edu/abs/1998MNRAS.298..525P}
}

@ARTICLE{2002MNRAS.329..897H,
       author = {{Hurley}, Jarrod R. and {Tout}, Christopher A. and {Pols}, Onno R.},
        title = "{Evolution of binary stars and the effect of tides on binary populations}",
      journal = {\mnras},
         year = 2002,
        month = feb,
       volume = {329},
       number = {4},
        pages = {897-928},
          doi = {10.1046/j.1365-8711.2002.05038.x},
archivePrefix = {arXiv},
       eprint = {astro-ph/0201220},
 primaryClass = {astro-ph},
       adsurl = {https://ui.adsabs.harvard.edu/abs/2002MNRAS.329..897H}
}

@ARTICLE{2001MNRAS.322..231K,
       author = {{Kroupa}, Pavel},
        title = "{On the variation of the initial mass function}",
      journal = {\mnras},
         year = 2001,
        month = apr,
       volume = {322},
       number = {2},
        pages = {231-246},
          doi = {10.1046/j.1365-8711.2001.04022.x},
archivePrefix = {arXiv},
       eprint = {astro-ph/0009005},
 primaryClass = {astro-ph},
       adsurl = {https://ui.adsabs.harvard.edu/abs/2001MNRAS.322..231K}
}

@ARTICLE{2017ApJS..230...15M,
       author = {{Moe}, Maxwell and {Di Stefano}, Rosanne},
        title = "{Mind Your Ps and Qs: The Interrelation between Period (P) and Mass-ratio (Q) Distributions of Binary Stars}",
      journal = {\apjs},
         year = 2017,
        month = jun,
       volume = {230},
       number = {2},
          eid = {15},
        pages = {15},
          doi = {10.3847/1538-4365/aa6fb6},
archivePrefix = {arXiv},
       eprint = {1606.05347},
 primaryClass = {astro-ph.SR},
       adsurl = {https://ui.adsabs.harvard.edu/abs/2017ApJS..230...15M}
}

@ARTICLE{1983ApJ...268..368E,
       author = {{Eggleton}, P.~P.},
        title = "{Aproximations to the radii of Roche lobes.}",
      journal = {\apj},
         year = 1983,
        month = may,
       volume = {268},
        pages = {368-369},
          doi = {10.1086/160960},
       adsurl = {https://ui.adsabs.harvard.edu/abs/1983ApJ...268..368E}
}

@ARTICLE{2022ApJ...925L..12K,
       author = {{Kupfer}, Thomas and {Bauer}, Evan B. and {van Roestel}, Jan and {Bellm}, Eric C. and {Bildsten}, Lars and {Fuller}, Jim and {Prince}, Thomas A. and {Heber}, Ulrich and {Geier}, Stephan and {Green}, Matthew J. and {Kulkarni}, Shrinivas R. and {Bloemen}, Steven and {Laher}, Russ R. and {Rusholme}, Ben and {Schneider}, David},
        title = "{Discovery of a Double-detonation Thermonuclear Supernova Progenitor}",
      journal = {\apjl},
         year = 2022,
        month = feb,
       volume = {925},
       number = {2},
          eid = {L12},
        pages = {L12},
          doi = {10.3847/2041-8213/ac48f1},
archivePrefix = {arXiv},
       eprint = {2110.11974},
 primaryClass = {astro-ph.SR},
       adsurl = {https://ui.adsabs.harvard.edu/abs/2022ApJ...925L..12K}
}

@ARTICLE{2012MNRAS.426.3282M,
       author = {{Maoz}, Dan and {Mannucci}, Filippo and {Brandt}, Timothy D.},
        title = "{The delay-time distribution of Type Ia supernovae from Sloan II}",
      journal = {\mnras},
         year = 2012,
        month = nov,
       volume = {426},
       number = {4},
        pages = {3282-3294},
          doi = {10.1111/j.1365-2966.2012.21871.x},
archivePrefix = {arXiv},
       eprint = {1206.0465},
 primaryClass = {astro-ph.CO},
       adsurl = {https://ui.adsabs.harvard.edu/abs/2012MNRAS.426.3282M}
}

@ARTICLE{2014MNRAS.444.3488F,
       author = {{Fuller}, Jim and {Lai}, Dong},
        title = "{Dynamical tides in compact white dwarf binaries: influence of rotation}",
      journal = {\mnras},
         year = 2014,
        month = nov,
       volume = {444},
       number = {4},
        pages = {3488-3500},
          doi = {10.1093/mnras/stu1698},
archivePrefix = {arXiv},
       eprint = {1406.2717},
 primaryClass = {astro-ph.SR},
       adsurl = {https://ui.adsabs.harvard.edu/abs/2014MNRAS.444.3488F}
}

@ARTICLE{2025A&A...693A.114B,
       author = {{Bhat}, Aakash and {Bauer}, Evan B. and {Pakmor}, R{\"u}diger and {Shen}, Ken J. and {Caiazzo}, Ilaria and {Rajamuthukumar}, Abinaya Swaruba and {El-Badry}, Kareem and {Kerzendorf}, Wolfgang E.},
        title = "{Supernova shocks cannot explain the inflated state of hypervelocity runaways from white dwarf binaries}",
      journal = {\aap},
         year = 2025,
        month = jan,
       volume = {693},
          eid = {A114},
        pages = {A114},
          doi = {10.1051/0004-6361/202451371},
archivePrefix = {arXiv},
       eprint = {2407.03424},
 primaryClass = {astro-ph.SR},
       adsurl = {https://ui.adsabs.harvard.edu/abs/2025A&A...693A.114B}
}

@ARTICLE{2025arXiv251012197B,
       author = {{Bhat}, Aakash and {Pakmor}, R{\"u}diger and {Shen}, Ken J. and {Bauer}, Evan B. and {Rajamuthukumar}, Abinaya Swaruba},
        title = "{Violent mergers can explain the inflated state of some of the fastest stars in the Galaxy}",
      journal = {arXiv e-prints},
         year = 2025,
        month = oct,
          eid = {arXiv:2510.12197},
        pages = {arXiv:2510.12197},
          doi = {10.48550/arXiv.2510.12197},
archivePrefix = {arXiv},
       eprint = {2510.12197},
 primaryClass = {astro-ph.SR},
       adsurl = {https://ui.adsabs.harvard.edu/abs/2025arXiv251012197B}
}

@ARTICLE{2025arXiv251011781P,
       author = {{Pakmor}, R{\"u}diger and {Shen}, Ken J. and {Bhat}, Aakash and {Rajamuthukumar}, Abinaya Swaruba and {Collins}, Christine E. and {O'Donnell}, Cillian and {Bauer}, Evan B. and {Callan}, Fionntan P. and {R{\"o}pke}, Friedrich K. and {Pollin}, Joshua M. and {Maguire}, Kate and {Kwok}, Lindsey A. and {Seth}, Ravi and {Taubenberger}, Stefan and {Justham}, Stephen},
        title = "{Violent mergers revisited: The origin of the fastest stars in the Galaxy}",
      journal = {arXiv e-prints},
         year = 2025,
        month = oct,
          eid = {arXiv:2510.11781},
        pages = {arXiv:2510.11781},
          doi = {10.48550/arXiv.2510.11781},
archivePrefix = {arXiv},
       eprint = {2510.11781},
 primaryClass = {astro-ph.HE},
       adsurl = {https://ui.adsabs.harvard.edu/abs/2025arXiv251011781P}
}

@ARTICLE{2019ApJ...887...68B,
       author = {{Bauer}, Evan B. and {White}, Christopher J. and {Bildsten}, Lars},
        title = "{Remnants of Subdwarf Helium Donor Stars Ejected from Close Binaries with Thermonuclear Supernovae}",
      journal = {\apj},
         year = 2019,
        month = dec,
       volume = {887},
       number = {1},
          eid = {68},
        pages = {68},
          doi = {10.3847/1538-4357/ab4ea4},
archivePrefix = {arXiv},
       eprint = {1906.08941},
 primaryClass = {astro-ph.SR},
       adsurl = {https://ui.adsabs.harvard.edu/abs/2019ApJ...887...68B}
}

@ARTICLE{2025ApJ...982....6S,
       author = {{Shen}, Ken J.},
        title = "{The Evolution of Hypervelocity Supernova Survivors and the Outcomes of Interacting Double White Dwarf Binaries}",
      journal = {\apj},
         year = 2025,
        month = mar,
       volume = {982},
       number = {1},
          eid = {6},
        pages = {6},
          doi = {10.3847/1538-4357/adb42e},
archivePrefix = {arXiv},
       eprint = {2502.04451},
 primaryClass = {astro-ph.SR},
       adsurl = {https://ui.adsabs.harvard.edu/abs/2025ApJ...982....6S}
}

@ARTICLE{2023ApJ...950L..10S,
       author = {{Shields}, Joshua V. and {Arunachalam}, Prasiddha and {Kerzendorf}, Wolfgang and {Hughes}, John P. and {Biriouk}, Sofia and {Monk}, Hayden and {Buchner}, Johannes},
        title = "{No Surviving SN Ia Companion in SNR 0509-67.5: Stellar Population Characterization and Comparison to Models}",
      journal = {\apjl},
         year = 2023,
        month = jun,
       volume = {950},
       number = {2},
          eid = {L10},
        pages = {L10},
          doi = {10.3847/2041-8213/acd6a0},
archivePrefix = {arXiv},
       eprint = {2305.03750},
 primaryClass = {astro-ph.SR},
       adsurl = {https://ui.adsabs.harvard.edu/abs/2023ApJ...950L..10S}
}

@ARTICLE{2025MNRAS.541.2231H,
       author = {{Hollands}, M.~A. and {Shen}, K.~J. and {Raddi}, R. and {G{\"a}nsicke}, B.~T. and {Bauer}, E.~B. and {Rebassa-Mansergas}, A.},
        title = "{Spectroscopic and kinematic analyses of a warm survivor of a D$^{6}$ supernova}",
      journal = {\mnras},
         year = 2025,
        month = aug,
       volume = {541},
       number = {3},
        pages = {2231-2245},
          doi = {10.1093/mnras/staf950},
archivePrefix = {arXiv},
       eprint = {2506.08081},
 primaryClass = {astro-ph.SR},
       adsurl = {https://ui.adsabs.harvard.edu/abs/2025MNRAS.541.2231H}
}

@ARTICLE{2024OJAp....7E...7B,
       author = {{Braudo}, Jessica and {Soker}, Noam},
        title = "{The runaway velocity of the white dwarf companion in the double detonation scenario of supernovae}",
      journal = {The Open Journal of Astrophysics},
         year = 2024,
        month = jan,
       volume = {7},
          eid = {7},
        pages = {7},
          doi = {10.21105/astro.2310.16554},
archivePrefix = {arXiv},
       eprint = {2310.16554},
 primaryClass = {astro-ph.HE},
       adsurl = {https://ui.adsabs.harvard.edu/abs/2024OJAp....7E...7B}
}

@ARTICLE{2024ApJ...975..127S,
       author = {{Shen}, Ken J. and {Boos}, Samuel J. and {Townsley}, Dean M.},
        title = "{Almost All Carbon/Oxygen White Dwarfs Can Host Double Detonations}",
      journal = {\apj},
         year = 2024,
        month = nov,
       volume = {975},
       number = {1},
          eid = {127},
        pages = {127},
          doi = {10.3847/1538-4357/ad7379},
archivePrefix = {arXiv},
       eprint = {2405.19417},
 primaryClass = {astro-ph.SR},
       adsurl = {https://ui.adsabs.harvard.edu/abs/2024ApJ...975..127S}
}

@ARTICLE{2024A&A...690A.368G,
       author = {{Geier}, S. and {Heber}, U. and {Irrgang}, A. and {Dorsch}, M. and {Bastian}, A. and {Neunteufel}, P. and {Kupfer}, T. and {Bloemen}, S. and {Kreuzer}, S. and {M{\"o}ller}, L. and {Schindewolf}, M. and {Schneider}, D. and {Ziegerer}, E. and {Pelisoli}, I. and {Schaffenroth}, V. and {Barlow}, B.~N. and {Raddi}, R. and {Geier}, S.~J. and {Reindl}, N. and {Rauch}, T. and {Nemeth}, P. and {G{\"a}nsicke}, B.~T.},
        title = "{A spectroscopic and kinematic survey of fast hot subdwarfs}",
      journal = {\aap},
         year = 2024,
        month = oct,
       volume = {690},
          eid = {A368},
        pages = {A368},
          doi = {10.1051/0004-6361/202450778},
archivePrefix = {arXiv},
       eprint = {2407.04479},
 primaryClass = {astro-ph.SR},
       adsurl = {https://ui.adsabs.harvard.edu/abs/2024A&A...690A.368G}
}

@ARTICLE{2022A&A...663A..91N,
       author = {{Neunteufel}, P. and {Preece}, H. and {Kruckow}, M. and {Geier}, S. and {Hamers}, A.~S. and {Justham}, S. and {Podsiadlowski}, Ph.},
        title = "{Properties and applications of a predicted population of runaway He-sdO/B stars ejected from single degenerate He-donor SNe}",
      journal = {\aap},
         year = 2022,
        month = jul,
       volume = {663},
          eid = {A91},
        pages = {A91},
          doi = {10.1051/0004-6361/202142864},
archivePrefix = {arXiv},
       eprint = {2112.07469},
 primaryClass = {astro-ph.SR},
       adsurl = {https://ui.adsabs.harvard.edu/abs/2022A&A...663A..91N}
}

@ARTICLE{2022MNRAS.512.6122C,
       author = {{Chandra}, Vedant and {Hwang}, Hsiang-Chih and {Zakamska}, Nadia L. and {Blouin}, Simon and {Swan}, Andrew and {Marsh}, Thomas R. and {Shen}, Ken J. and {G{\"a}nsicke}, Boris T. and {Hermes}, J.~J. and {Putterman}, Odelia and {Bauer}, Evan B. and {Petrosky}, Evan and {Dhillon}, Vikram S. and {Littlefair}, Stuart P. and {Ashley}, Richard P.},
        title = "{The SN Ia runaway LP 398-9: detection of circumstellar material and surface rotation}",
      journal = {\mnras},
         year = 2022,
        month = jun,
       volume = {512},
       number = {4},
        pages = {6122-6133},
          doi = {10.1093/mnras/stac883},
archivePrefix = {arXiv},
       eprint = {2110.06935},
 primaryClass = {astro-ph.SR},
       adsurl = {https://ui.adsabs.harvard.edu/abs/2022MNRAS.512.6122C}
}

@ARTICLE{2024MNRAS.527.2072D,
       author = {{Deshmukh}, Kunal and {Bauer}, Evan B. and {Kupfer}, Thomas and {Dorsch}, Matti},
        title = "{Modelling the AM CVn and double detonation supernova progenitor binary system CD-30{\textdegree}11223}",
      journal = {\mnras},
         year = 2024,
        month = jan,
       volume = {527},
       number = {2},
        pages = {2072-2082},
          doi = {10.1093/mnras/stad3288},
archivePrefix = {arXiv},
       eprint = {2310.01293},
 primaryClass = {astro-ph.SR},
       adsurl = {https://ui.adsabs.harvard.edu/abs/2024MNRAS.527.2072D}
}

@ARTICLE{2021A&A...646L...8N,
       author = {{Neunteufel}, P. and {Kruckow}, M. and {Geier}, S. and {Hamers}, A.~S.},
        title = "{Predicted spatial and velocity distributions of ejected companion stars of helium accretion-induced thermonuclear supernovae}",
      journal = {\aap},
         year = 2021,
        month = feb,
       volume = {646},
          eid = {L8},
        pages = {L8},
          doi = {10.1051/0004-6361/202040022},
archivePrefix = {arXiv},
       eprint = {2011.14887},
 primaryClass = {astro-ph.SR},
       adsurl = {https://ui.adsabs.harvard.edu/abs/2021A&A...646L...8N}
}

@ARTICLE{2020A&A...641A..52N,
       author = {{Neunteufel}, P.},
        title = "{Exploring velocity limits in the thermonuclear supernova ejection scenario for hypervelocity stars and the origin of US 708}",
      journal = {\aap},
         year = 2020,
        month = sep,
       volume = {641},
          eid = {A52},
        pages = {A52},
          doi = {10.1051/0004-6361/202037792},
archivePrefix = {arXiv},
       eprint = {2006.11427},
 primaryClass = {astro-ph.SR},
       adsurl = {https://ui.adsabs.harvard.edu/abs/2020A&A...641A..52N}
}

@ARTICLE{2009A&A...493.1081J,
       author = {{Justham}, S. and {Wolf}, C. and {Podsiadlowski}, Ph. and {Han}, Zh.},
        title = "{Type Ia supernovae and the formation of single low-mass white dwarfs}",
      journal = {\aap},
         year = 2009,
        month = jan,
       volume = {493},
       number = {3},
        pages = {1081-1091},
          doi = {10.1051/0004-6361:200810106},
archivePrefix = {arXiv},
       eprint = {0811.2633},
 primaryClass = {astro-ph},
       adsurl = {https://ui.adsabs.harvard.edu/abs/2009A&A...493.1081J}
}

\end{document}